\documentclass[fleqn,usenatbib]{mnras}

\usepackage{newtxtext,newtxmath}
\usepackage[T1]{fontenc}

\DeclareRobustCommand{\VAN}[3]{#2}
\let\VANthebibliography\thebibliography
\def\thebibliography{\DeclareRobustCommand{\VAN}[3]{##3}\VANthebibliography}

\usepackage{graphicx}	% Including figure files
\usepackage{amsmath}	% Advanced maths commands
\usepackage{multicol}
\usepackage{bm}
\usepackage{cancel}
\title[Inclination Instability]{Inclination instability of circumbinary planets }

\author[Stephen H. Lubow et al.]{
Stephen H. Lubow$^{1}$\thanks{E-mail: lubow@stsci.edu},
Anna C. Childs$^{2}$
and
Rebecca G. Martin$^{3,4}$
\\
$^1$Space Telescope Science Institute, 3700 San Martin Drive, Baltimore, MD 21218, USA\\
$^{2}$Center for Interdisciplinary Exploration and Research in Astrophysics (CIERA) and Department of Physics and Astronomy Northwestern University, \\
1800 Sherman Ave, Evanston, IL 60201 USA\\
$^3$Nevada Center for Astrophysics, University of Nevada, Las Vegas, 4505 S. Maryland Pkwy., Las Vegas, NV 89154, USA\\
$^4$Department of Physics and Astronomy,University of Nevada, Las Vegas, 4505 S. Maryland Pkwy., Las Vegas, NV 89154, USA\\
}

\date{Accepted April 9, 2024. Received March 19, 2024; in original form January 24, 2024}

\pubyear{2024}

\begin{document}
\label{firstpage}
\pagerange{\pageref{firstpage}--\pageref{lastpage}}
\maketitle

\begin{abstract}
We analyze  a  tilt instability of the orbit of an outer planet in a two planet circumbinary system that we recently reported.
The binary is on an eccentric orbit and the inner circumbinary planet is on a circular polar orbit that causes the the binary to undergo  apsidal precession.
 The outer circumbinary planet is initially on a circular or eccentric orbit that is coplanar with respect to the binary. 
 We apply a Hamiltonian in quadrupole order of the binary potential to show that the tilt instability is the result of a secular resonance in which the apsidal precession rate of the binary matches the nodal
 precession  rate of the outer planet. Resonance is possible because the polar inner planet causes the apsidal precession of the binary to be retrograde.
 The outer planet periodically undergoes large tilt oscillations for which we analytically determine the initial evolution and maximum inclination.
 Following a typically relatively short adjustment phase, the tilt grows exponentially in time at a characteristic rate that is of order the absolute value of the binary apsidal precession rate.
 The  analytic results agree well with numerical simulations. This instability is analogous to the Kozai-Lidov instability, but applied to a circumbinary object.
 The instability fails to operate if the binary mass ratio is too extreme.
 The instability occurs even if the outer planet  is instead an object of stellar mass and involves tilt oscillations of the inner binary.
%These results may help to explain the lack of observed circumbinary planets and can aid future observations in the detection of highly inclined planets.
\end{abstract}


%%%%%%%%%%%%%%%%%%%%%%%%%%%%%%%%%%%%%%%%%%%%%%%%%%

%%%%%%%%%%%%%%%%% BODY OF PAPER %%%%%%%%%%%%%%%%%%

\section{Introduction}
About a dozen circumbinary planets have been detected through transits with the Kepler and TESS telescopes.
All of these planets are nearly coplanar with the orbit of the binary. This is largely
a selection effect due to the technique adopted \citep{Schneider,MartinD2014,MartinD2015,MartinD2017,Zhang2019}. Estimates suggest that the frequency of %\cancel{\ACC{occurrence}} 
these observed circumbinary
planets is similar to that around single stars, once the selection effects are taken into account \citep{Li2016}. On this basis one might
conclude that there is not a large population of noncoplanar circumbinary planets. 

However, misaligned planets
are more likely to be found around longer period binaries than the planets found in current observations.
Circumbinary planets are expected to form in circumbinary discs. Circumbinary discs are often
found in observations to be misaligned \citep{Czekala2019}. If the binary orbit is sufficiently eccentric, the protostellar disc
could increase its inclination and evolve to a polar state in which it is perpendicular to the binary
orbital plane \citep{Aly2015, Martin17,Lubow2018,Zanazzi2018,Cuello2019,Smallwood2020}.  Due to tidal dissipation, closer binaries, with periods less than about 10 days,
tend to have low eccentricity \citep{Goldman1991, Raghavan2010}, which is unfavorable for the growth of disc inclination. Circumbinary
discs around low eccentricity binaries typically evolve to coplanarity \citep{Nixon2011,Facchini2013,Foucart2014}. Most of the detected circumbinary
planets are found around such lower eccentricity binaries. At longer binary orbital periods, circumbinary
discs are more likely to be misaligned with respect to the binary, since the binary eccentricities are typically larger.
Observations suggest that circumbinary discs in orbit around binaries with periods longer than about 30 days
tend to be misaligned with respect to the binary \citep{Czekala2019}. In addition,  two polar gas discs and one polar debris disc have been detected and both
involve highly eccentric binaries ($e_{\rm } \sim 0.8$) \citep{Kennedy2012,Kennedy2019,kenworthy2022}.

Since there may be a population of  misaligned circumbinary planets, some recent studies have examined their
predicted properties \citep[e.g.,][]{Verrier2009,Farago2010,Doolin2011,Naoz2017,Quarles2018,Chen2019}.  The orbit of a slightly misaligned planet around an eccentric orbit binary undergoes circulation
in which its line of nodes pass through all $360^{\circ}$ and the angular momentum vector precesses around the binary
angular momentum vector.  Because the binary is eccentric, the orbit of the planet undergoes tilt oscillations.
But at higher  levels of initial misalignment, the orbit of the planet can undergo libration in which its line of nodes 
passes through a limited angular range, less than $360^{\circ}$
In this case, the planet's  
angular momentum vector precesses
around the binary eccentricity vector, rather than around the binary angular momentum vector.
\cite{Chen2019} analyzed the properties of these orbits as a function of planet mass. \cite{Chen2020} studied the stability of noncoplanar circumbinary planets and found that the polar configuration is the most stable at high binary eccentricity.

More recently, some studies have concentrated on the form of planet orbits and their stability when
there are two interacting planets \citep[e.g.][]{Chen2023ffp}. \cite{Chen2022} considered a configuration in which both planet orbits are initially
mutually coplanar, but misaligned relative to the orbit plane of the binary.
Planet-planet interactions can lead to complex tilt oscillations of each
planet because two torques operate at independent frequencies: the nodal precession frequency due to the binary
and the nodal precession frequency due to the relative nodal precession of the two planets.
In addition, circumbinary planet-planet interactions can make the planet orbit much less stable than would
occur for a single circumbinary planet or two interacting planets around a single star \citep{Chen2023}.

Recently \cite{Childs2023} investigated the orbital evolution of a two planet circumbinary system in which
the planets are mutually initially highly misaligned. The inner planet is on a polar orbit, while the outer planet
is coplanar with the orbit of the binary. Such a configuration could develop from planet formation in  a broken disk \citep{Nixon2013,Facchini2013,Lubow2018,Martin2018,Martin2019} or,
from multiple epochs of disc formation \citep{Bate2018}.
The inner planet is assumed to be sufficiently far from the outer planet
that the direct planet-planet interactions are small compared to the effects of the binary on the outer planet.
The surprising result is that the outer planet undergoes large tilt oscillations away from the coplanar
configuration. Some analytic estimates were made for the radial range over which this effect operates.
The purpose of this paper is to further explore the dynamics of this configuration. 

In Section \ref{sec:ana} we describe an analytic model for the
orbital evolution of the outer planet. In Section \ref{sec:geom}
we describe a geometric explanation for qualitative features 
of the analytic model. Section \ref{sec:num} describes some results
of simulations and makes a comparison with  the predictions
of the analytic model. It discusses how the instability operates if the
binary mass ratio is extreme or if the outer object is massive. Sections \ref{sec:disc} and \ref{sec: sum}
contain the discussion and summary, respectively.

\section{Analytic Model}
\label{sec:ana}
We consider a binary star system of mass $m_{\rm b}$ with component masses $m_1$ and $m_2$
that is on an orbit with semi-major axis $a_{\rm b}$, eccentricity $e_{\rm b}$, and orbital frequency $\Omega_{\rm b}$.
The binary undergoes apsidal precession at a rate $d \varpi_{\rm b}/dt$, where $\varpi_{\rm b}$ is the longitude of periapsis of the binary.
 We consider two circumbinary planets.
The inner planet with mass $m_3$ is on a circular polar orbit with semi-major $a_3$. The outer planet
is modeled as a test particle, $m_4=0$. 
 The outer planet has orbital elements that are denoted with the notation $a_4, e_4, i, \omega_4,$ and $\Omega_4$ for its
semi-major axis, eccentricity,  inclination with respect to the binary orbital plane, argument of periapsis, and longitude of ascending node in the inertial frame, respectively.  The orbit may have have an initial eccentricity and is  initially nearly coplanar with respect to the binary.

As shown in \cite{Childs2023}, the gravitational forcing of the outer planet by the inner planet can typically be ignored compared with 
the forcing by the binary, provided that the two planets are well separated.  
The inner planet causes the binary to undergo apsidal precession.  The Hamiltonian that describes the motion of the outer planet then
depends on the binary parameters and its precession rate.  We apply a Hamiltonian to quadrupole order
in the binary potential. %We consider cases in which the inner planet is on a circular orbit and the outer planet is on a circular or eccentric orbit.  
We note that in  the case that  the outer planet is on a circular orbit or the binary has equal mass members,  the octupole order terms vanish \citep[e.g., equation (7) of][]{Elia2019} and the quadrupole approximation is expected to be more accurate.

\subsection{Hamiltonian}\label{sec:Hamiltonian}

A  secular  Hamiltonian
is obtained by averaging the potential due to the binary over a binary orbit period and  averaging that potential on the outer planet over its orbital period.
The secular equations of motion for the orbital elements of the outer planet are derived
from the secular Hamiltonian.
In the absence of the inner planet, the binary orbit is fixed in the inertial frame. 
The unperturbed Hamiltonian is obtained by considering the binary to be a single point mass.
The equations of motion for the outer planet are derived from the perturbed Hamiltonian that 
is  due to tidal effects of the binary.
The secular perturbed Hamiltonian per unit mass of the outer planet for the nonprecessing binary (np) to quadrupole order
in the binary potential is given by
\begin{equation}
H_{\rm np} = \frac{\alpha}{(1-e_4^2)^{3/2} } \left[ (2+3e_{\rm b}^2)(1-3 \cos^2{(i_4)}) - 15 e_{\rm b}^2 \cos{(2 \Omega_4)} \sin^2{i_4} \right],
  \label{H} 
\end{equation}
where
\begin{equation}
\alpha = \frac{1}{16} \frac{m_1 m_2}{m_{\rm b}^2} \frac{a_{\rm b}^5}{a_4^3} \Omega_{\rm b}^2 \label{alpha}
\end{equation}
\citep[e.g.,][]{Farago2010, Naoz2017} and the binary angular frequency is $\Omega_{\rm b}=\sqrt{G m_{\rm b}/a_{\rm b}^3}$. 
This Hamiltonian is expressed in terms of orbital elements. Appendix A describes the Hamiltonian with Delaunay 
canonical variables.

In the presence of the inner planet, the binary undergoes apsidal precession at rate $d \varpi_{\rm b}/dt$.
The orbit of the outer planet is best described in the frame that precesses with the binary \citep[e.g.,][]{Farago2010, Zanardi2018,Zanardi2023}.
In that frame, the inclination is a single valued function of nodal phase (modulo $360^\circ$) 
over all times. Consequently, we analyze the motion of the outer planet in the frame that precesses with the binary.
 In transforming to this frame, we replace $\Omega_4$ 
by the value of the longitude of the ascending node in the rotating frame, denoted as $\phi_4$, while inclination $i_4$ remains unchanged from its value in the nonrotating frame. The transformation is described in more detail using Delaunay canonical variables in Appendix A.
The Hamiltonian $H_{\rm np}$ is  transformed
to a Hamiltonian that accounts for the frame rotation by  adding a term  \citep[see Appendix D of][]{Tremaine2023}.
 The added term is $- {\bm \Omega}_{\rm f} \cdot {\bm L_4}$, where  ${\bm \Omega}_{\rm f}$  is the rotation rate of the frame  and ${\bm L_4}$
is the angular momentum per unit mass of the outer planet 
with magnitude
\begin{equation}
L_4=  a_{\rm b}^{3/2} \Omega_{\rm b} \sqrt{a_4 (1-e_4^2)}.
\end{equation}
In a frame that rotates with the binary eccentricity vector,  the Hamiltonian becomes
\begin{equation}
\begin{aligned}
H = \frac{\alpha}{(1-e_4^2)^{3/2} } & \left[ (2+3e_{\rm b}^2)(1-3 \cos^2{(i_4)}) - 15 e_{\rm b}^2 \cos{(2 \phi_4)} \sin^2{i_4} \right] \label{Ht}  \\
     &-L_4 \frac{ d\varpi_{\rm b}}{dt}  \cos{(i_4)}.
 \end{aligned}
 \end{equation}
%In this equation 
%\begin{eqnarray}
%H &=& \frac{\alpha}{(1-e_4^2)^{3/2} } \left[ (2+3e_{\rm b}^2)(1-3 \cos^2{(i)}) - 15 e_{\rm b}^2 \cos{(2 \phi)} \sin^2{i} \right] \label{Ht}  \\
%    & & -a_{\rm b}^{3/2} a^{1/2} \Omega_{\rm b} \frac{ d\varpi_{\rm b}}{dt}  \cos{(i)} \nonumber.
 %\end{eqnarray}
In this equation $\phi_4$ is the longitude of the ascending node of the outer planet in the corotating frame, i.e., relative to the instantaneous eccentricity vector of the binary.
The apsidal precession rate of the binary due to the polar planet is given by
\begin{equation}
\frac{d \varpi_{\rm b}}{dt} =  - \frac{9}{4}  \, \frac{m_3} {m_{\rm b}} \, \left( \frac{a_{\rm b}} {a_3} \right)^3 
%\frac{\sqrt{1-e_{\rm b}^2}}{(1-e_3^2)^{3/2}} \, \Omega_{\rm b} 
\sqrt{1-e_{\rm b}^2} \, \Omega_{\rm b}
\label{dvarpidt}
 \end{equation}
 \citep[e.g.,][]{Innanen1997,Naoz2016,Zhang2019, Childs2023}.

\subsection{Equations of Motion}

We apply Hamilton's  equations with Delanuay variables  to the Hamiltonian given by Equation (\ref{Ht}) (see Appendix A)
 to obtain
\begin{eqnarray}
\frac{d a_4}{dt} &=& 0, \label{dadt}\\
\frac{d e_4}{dt} &=& 0,  \label{dedt} \\
\frac{d i_4}{dt} &=& \frac{1}{L_4 \sin{(i_4)} } \frac{\partial H}{\partial \phi_4},   \label{didt0}\\
\frac{d \phi_4}{dt} &=&  -\frac{1}{L_4 \sin{(i_4)} } \frac{\partial H}{\partial i_4}.   \label{dphidt0}
\end{eqnarray}
Consequently, the  semi-major axis and eccentricity remain constant
 during the orbital evolution.
We then obtain
\begin{eqnarray}
\frac{ d i_4}{dt}  &=& \frac{15} {8} \frac{\beta}{(1-e_4^2)^2} \ e_{\rm b}^2 \sin{i_4} \sin{2 \phi_4} \label{didt} \\
\frac{ d \phi_4}{dt}  &=& -\frac{3} {8} \frac{\beta}{(1-e_4^2)^2} \cos{(i_4)} \, ( 2 + 3 e_{\rm b}^2- 5 e_{\rm b}^2 \cos{(2 \phi_4)} ) \label{dphidt} \nonumber \\
& &  - \frac{d \varpi_{\rm b}}{dt}, 
%\frac{ d \varpi}{dt}  &=& - 3 \omega_{\rm a}  \sqrt{1-e_{\rm b}^2}  \label{dvarpidt},
\end{eqnarray}
where $d \varpi_{\rm b}/dt$ is given by Equation (\ref{dvarpidt}) and 
\begin{eqnarray}
\beta  &=& \frac{m_1 m_2}{m_{\rm b}^2} \Omega_{\rm b}  \left( \frac{a_{\rm b} }{a_4} \right)^{7/2}% \\
\end{eqnarray}
%is a constant in time that 
is constant in time and is related to the magnitude of the nodal precession rate  of the outer planet.
Somewhat similar equations have been derived by \cite{Zanardi2018} in another context.

\subsection{Resonance Condition}
Equation (\ref{dphidt}) has a simple physical interpretation. It can be written as the sum of two contributions 
\begin{equation}
\frac{ d \phi_4}{dt} = \frac{ d \Omega_{\rm inertial}}{dt} + \frac{ d \Omega_{\rm rot}} {dt}. \label{phisum}
\end{equation}
The first term on the right hand side, $d \Omega_{\rm inertial}/dt$, is the nodal precession rate of the outer planet  due to the binary in the inertial frame,
 while the second term, $d \Omega_{\rm rot}/dt$, is the nodal precession rate of the outer planet in the corotating frame due to the rotation of the reference frame. 
 From the geometry of inclined orbits in the rotating frame, it follows that $d \Omega_{\rm rot}/dt = - d \varpi_{\rm b}/dt$.
 %These two contributions then respectively correspond to the two terms on the RHS of Equation (\ref{dphidt}).
 
 For small inclination, $i_4 \simeq 0$, a resonance is possible when $d \phi_4/dt=0$ so that the outer planet's orbit evolves at constant nodal phase in the frame
 of the binary. From Equation (\ref{dphidt}), this is possible when the nodal precession rate of the particle in the inertial frame,
 the first term on the RHS of Equation (\ref{phisum}) matches the apsidal precession rate of the binary $d \varpi_{\rm b}/dt$.  Since both are negative,
 resonance is possible.  For a fixed set of the binary parameters
 $a_{\rm b}$ and $e_{\rm b}$, we then expect that the resonance  condition is satisfied for a range of the
 semi-major axis of the outer planet $a_4$, since $\cos{(2 \phi_4)}$ can take on values between -1 and +1. For values of $\phi_4$ for
 which the RHS of Equation  (\ref{dphidt}) vanishes, the particle is locked at that phase since  $d \phi_4/dt=0$.
 The resonance condition then requires that
 \begin{equation}
 a_{\rm i} < a_4 < a_{\rm o}, \label{acond}
 \end{equation}
 where 
 \begin{equation}
  a_{\rm i} = a_{\rm b} \left(  \left(   \frac{a_3} {a_{\rm b} }  \right)^3 
  %\frac{(1-e_3^2)^{3/2} \sqrt{1-e_{\rm b}^2 }}{ 3 (1-e_4^2)^2}\, \frac{m_1 m_2}{m_{\rm b} m_3} \right)^{2/7} \label{ai}
\frac{ \sqrt{1-e_{\rm b}^2 }}{ 3 (1-e_4^2)^2}\, \frac{m_1 m_2}{m_{\rm b} m_3} \right)^{2/7} \label{ai}
  \end{equation}
  and
 \begin{equation}
  a_{\rm o} = a_{\rm b} \left(  \left(   \frac{a_3} {a_{\rm b} }  \right)^3 
   \left(\
   \frac{ (1+4 e_{\rm b}^2)}{ 3 (1-e_4^2)^2 \sqrt{1-e_{\rm b}^2} } 
   \right) 
   \frac{m_1 m_2}{m_{\rm b} m_3} \right)^{2/7} \label{ao}.
  \end{equation}
  The dimensionless ratio of inner to outer radii is given by 
  \begin{equation}
  \frac{ a_{\rm i}}{a_{\rm o}} = \left( \frac{1-e_{\rm b}^2}{1+4e_{\rm b}^2} \right)^{2/7} .\label{width}
  \end{equation}
This ratio increases with increasing binary eccentricity and is independent of the inner and outer planet eccentricities.
  For the case of a circular orbit   outer planet, these
  critical radii reduce to equations~(9) and~(10) in \cite{Childs2023}. 
  The motion of the outer planet can undergo libration involving angle $\phi_4$, as we will see in the phase portrait in Figure \ref{fig:phpl}. 
  The orbit that satisfies the resonance condition is the largest  librating orbit in a phase portrait and passes through the origin.
   Libration is also seen in Figure 2 of \cite{Childs2023}. 
  %They were based on the requirement that orbits that start at zero inclination be librating  in the frame of the binary. Here we give the interpretation in terms of a resonance condition.

\subsection{Maximum Inclination}
We use the fact that  $H$ given by Equation (\ref{Ht}) is a constant of motion to determine the maximum inclination $i_{\rm max}$ of the initially coplanar outer planet.
For an initially coplanar orbit, we have that 
\begin{equation}
H_{\rm i0} = - \frac{ 2 \alpha}{(1-e_4^2)^{3/2} }  (2+3e_{\rm b}^2)  
  -L_4 \frac{ d\varpi_{\rm b}}{dt}.   \label{H0} 
\end{equation}
For librating orbits in the frame of the binary, the maximum inclination occurs for $di_4/dt=0$ in Equation (\ref{didt}) and is for $\phi_4=90^{\circ}$, also as seen in Fig. 2 of \cite{Childs2023}. In that case, $H$ is given by
\begin{align}
H_{\rm imax} = \frac{\alpha}{(1-e_4^2)^{3/2} }  \left[ (2+3e_{\rm b}^2)(1-3 \cos^2{(i_{\rm max})}) 
    + 15 e_{\rm b}^2  \sin^2{(i_{\rm max})} \right] \nonumber \\
    -L_4 \frac{ d\varpi_{\rm b}}{dt} \cos{(i_{\rm max})}  \label{H90}.
 \end{align}
 Setting $H_{\rm i0} =H_{\rm imax}$ we obtain
 \begin{equation}
  i_{\rm max} =  \arccos{\left( -1 + 2 \,x^{7/2} \right)}, \label{imax}
\end{equation}
where
\begin{equation}
x= \frac{a_4}{a_{\rm o}}
\end{equation}
and $a_{\rm o}$ is given by Equation (\ref{ao}).  

Equation (\ref{imax}) holds only for  
orbits at resonance, $a_{\rm i} < a_4 < a_{\rm o}$, because such orbits undergo libration.
Thus,  it holds for $[(1-e_{\rm b}^2)/(1+4 e_{\rm b}^2))]^{2/7} < x <1$.
Notice that Equation (\ref{imax}) implies that $i_{\rm max}$ decreases with increasing $a_4$ and approaches zero
as $a_4$ approaches $a_{\rm o}$. 

%\RGM{Should figure 6 be moved and discussed here first?}

\subsection{Inclination Instability}
We show that the outer planet's inclination grows as a result of an instability.
For $i_4$ exactly equal to zero, Equation (\ref{didt}) implies that $i_4$ remains at zero. But this state is unstable.
 We consider the case  that the initial inclination $i_4(0)=i_0$ is nonzero  and small.
For $a_{\rm i} < a_4 < a_{\rm o}$, Equation (\ref{dphidt})  in lowest order is independent of $i_4$ 
and admits a solution for $\phi_4$ which is constant
in time. We denote that solution as $\phi_{\rm c}$ and restrict the definition of $\phi_{\rm c}$ to range from $0$ to $90^\circ$. It is given by
\begin{equation}
\phi_{\rm c} = \frac{1}{2} \arccos{\left ( \frac{2 + 3  e_{\rm b}^2  - 2 x^{7/2} (1 + 4 e_{\rm b}^2)}{5 e_{\rm b}^2  }\right)}.
\label{phic}
\end{equation}
 It follows that  $\phi_c=0$ for $a_4=a_{\rm i}$ and $\phi_c=90^\circ$ for $a_4=a_{\rm o}$.
 Using Equation (\ref{didt}), we find that the growth rate $\gamma_{\rm c}=d \ln{i} /dt$ of the instability for small $i_4$  and $\phi_4=\phi_{\rm c}$ is constant in time. It is given by
 %\begin{equation}
 %\gamma_{\rm c} =  \frac{3}{4}  \frac{\beta}{(1-e_4^2)^2} \sqrt{  (1 + 4 e_{\rm b}^2)
 %(1 - x^{7/2}) ( x^{7/2} (1 + 4 e_{\rm b}^2) -(1 -e_{\rm b}^2))}. \label{gammac}
 % \end{equation}
 \begin{equation}
 \gamma_{\rm c} =     \frac{\lambda}{x^{7/2}} \sqrt{ 
 (1 - x^{7/2}) ( x^{7/2} (1 + 4 e_{\rm b}^2) -(1 -e_{\rm b}^2))}, \label{gammac}
  \end{equation}
  where
\begin{equation}
\lambda=\frac{9}{4}   \left( \frac{a_{\rm b}}{a_3} \right)^3 \left( \frac{m_3}{m_{\rm b}} \right)
\sqrt{\frac{1-e_{\rm b}^2}{1+4e_{\rm b}^2}} %\frac{\Omega_{\rm b}}{(1-  e_3^2)^{3/2}}.
\Omega_{\rm b}.
\end{equation}

 It follows that $\gamma_{\rm c} =0$ at $a_4=a_{\rm i}$ and $a_4=a_{\rm o}$ and is 
  positive for $a_{\rm i} < a_4 < a_{\rm o}$. For $\phi_4=\phi_{\rm c}$, the inclination therefore grows exponentially. The growth rate is of order $\lambda$ that is of order the binary apsidal precession rate $|d \varpi_{\rm b}/dt|$  or faster provided that $e_{\rm b}$ is not small, as  is shown below.  
  %that is of order the nodal precession period  of the outer planet in the inertial frame that is also the apsidal precession period of the binary.
  
\begin{figure}
	\includegraphics[width=\columnwidth]{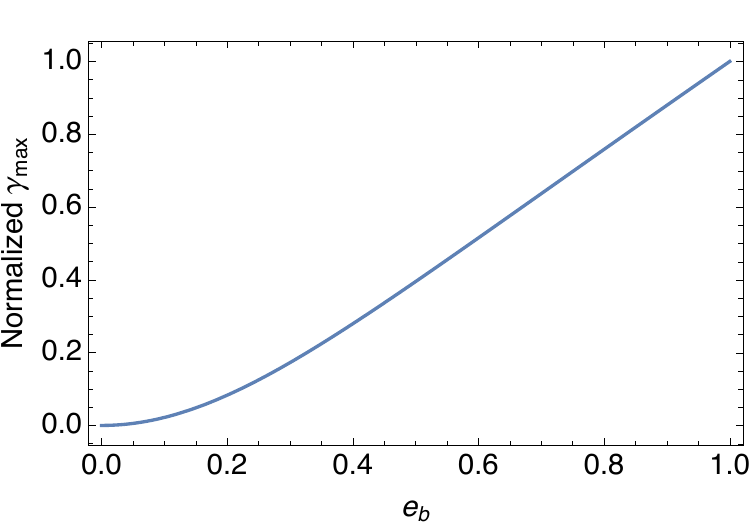}
    \caption{Normalized maximum growth rate of the inclination of an initially nearly coplanar outer planet as a function of binary eccentricity.
The maximum is taken over the possible values of the planet semi-major axis with all other parameters fixed. The normalization
factor is 
such that the plotted function is unity at $e_{\rm b}=1$ (see Equation (\ref{gammam})).    }
    \label{fig:grm}
\end{figure}

  We consider a fixed set of system parameters but allow the outer planet semi-major axis to vary. The growth rate $\gamma_{\rm c}$
  achieves a maximum value for an intermediate  value of $a_4$ denoted by $a_{\rm m}$ that lies between $a_{\rm i}$ and $a_{\rm o}$. It is given by
  \begin{equation}
  a_{\rm m} = \left ( \frac{1-e_{\rm b}^2}{1+ 3/2  \, e_{\rm b}^2} \right)^{2/7} a_{\rm o}. \label{am}
  \end{equation}
  The maximum growth rate is given by
  \begin{equation}
  \gamma_{\rm max} = \frac{45}{8}  \left( \frac{a_{\rm b} }{a_3} \right)^3 \frac{m_3}{m_{\rm b}} \frac{e_{\rm b}^2}{\sqrt{1+4 e_{\rm b}^2}} %\frac{\Omega_{\rm b}}{(1-  e_3^2)^{3/2}} . \label{gammam}
  \Omega_{\rm b}. \label{gammam}
  \end{equation}
  Notice that $\gamma_{\rm max}$ is independent of the outer planet's eccentricity $e$ and the binary mass ratio. We note, however, that the binary mass ratio cannot be too extreme. Otherwise,
  the 
  approximation that the direct interaction of the inner planet with the outer planet
  is much less important than tidal interaction of the binary with the outer planet
  breaks down, as is explored in Section \ref{sec:q}. %In addition, if $a_{\rm m}$ is small enough as occurs at very high values of $e_{\rm b}$, the quadrupole approximation breaks down. 
  For small binary eccentricity, this growth rate is quadratic in $e_{\rm b}$, but varies almost linearly in $e_{\rm b}$ for larger binary eccentricity
  (see Figure~\ref{fig:grm}).
  The change of inclination over a nodal precession period of the binary is given by 
  \begin{equation}
   \Delta \ln{i} = \gamma_{\rm max} \frac{2 \pi}{|d \varpi_{\rm b}/dt|} =\frac{  5 \pi e_{\rm b}^2 }
   {\sqrt{(1-e_{\rm b}^2)(1+4 e_{\rm b}^2)}}.
  \end{equation}
  Therefore a substantial growth of inclination can occur over an apsidal period of the binary provided that $e_{\rm b}$
  is not small. It is large for high binary eccentricity.

\subsection{General Initial Conditions}
\label{sec:ic}
Equation (\ref{gammac}) shows that inclination growth occurs exponentially fast for  $a_{\rm i} < a_4 < a_{\rm o}$, until $i_4$ becomes of order unity,
provided that the initial nodal phase $\phi_0$ equals $\phi_c$. We explore the evolution of inclination for arbitrary
initial phases  $\phi_0$, again with small initial $i_4$.
For  $0<\phi_0< 90^\circ$, Equation (\ref{dphidt}) has a solution
\begin{equation}
\phi_4(t) = \arctan{\left(\tan{(\phi_{\rm c})} \frac{u_1(t)}{ u_2(t)} \right)}, \label{phit}
\end{equation}
where
\begin{eqnarray}
u_1(t) &=& \tan{(\phi_0)} + \tan{(\phi_{\rm c})} \tanh{(\gamma_{\rm c} t)}, \label{u1}\\
u_2(t) &=& \tan{(\phi_{\rm c})} + \tan{(\phi_0)} \tanh{(\gamma_{\rm c} t)}. \label{u2}
\end{eqnarray}
On secular timescales, $\gamma_{\rm c} t \ga 1$, $u_1$ approaches $u_2$ and so $\phi_4(t)$ approaches $\phi_{\rm c}$ by Equation (\ref{phit}),
Other branches of the solution that occur outside this range of $\phi_0$ can be constructed, but we do not show them.

For small $i_4$  and any $\phi_0$, Equations (\ref{didt}) and (\ref{dphidt}) have a solution for the instantaneous early growth rate $\gamma(t) = d\ln{i}/dt$ that is given by 
%(t1v t2v)/(t2v^2 Cos[phica]^2 + t1v^2 Sin[phica]^2)
\begin{equation}
\gamma(t) = \frac{   u_1(t) u_2(t) \gamma_{\rm c}}{u_1(t)^2 \sin^2{ \phi_{\rm c} }  + u_2(t)^2 \cos^2{ \phi_{\rm c} }}  
\label{gamma} .
\end{equation}

For $0<\phi_0 < 90^\circ$ and  $180^\circ<\phi_0 < 270^\circ$, we have that $\sin{(2 \phi_0)} >0 $ and so by Equation (\ref{didt})  the inclination initially grows, $\gamma(0)>0$.
In fact from Equation (\ref{gamma}) it follows that $\gamma(t)>0$ as long as $i_4$ remains small.
In that case, from Equations (\ref{u1}) and (\ref{u2}), it follows that $u_1(t)$ approaches $u_2(t)$ on a timescale of $\gamma_{\rm c}^{-1}$.
This timescale is relatively short because many timescales  $\gamma_{\rm c}^{-1}$ are required for the initially small inclination to grow to order unity values.
Beyond that timescale, $\gamma(t)$ approaches $\gamma_{\rm c}$ and the growth becomes exponential in time.

\begin{figure}
	\includegraphics[width=\columnwidth]{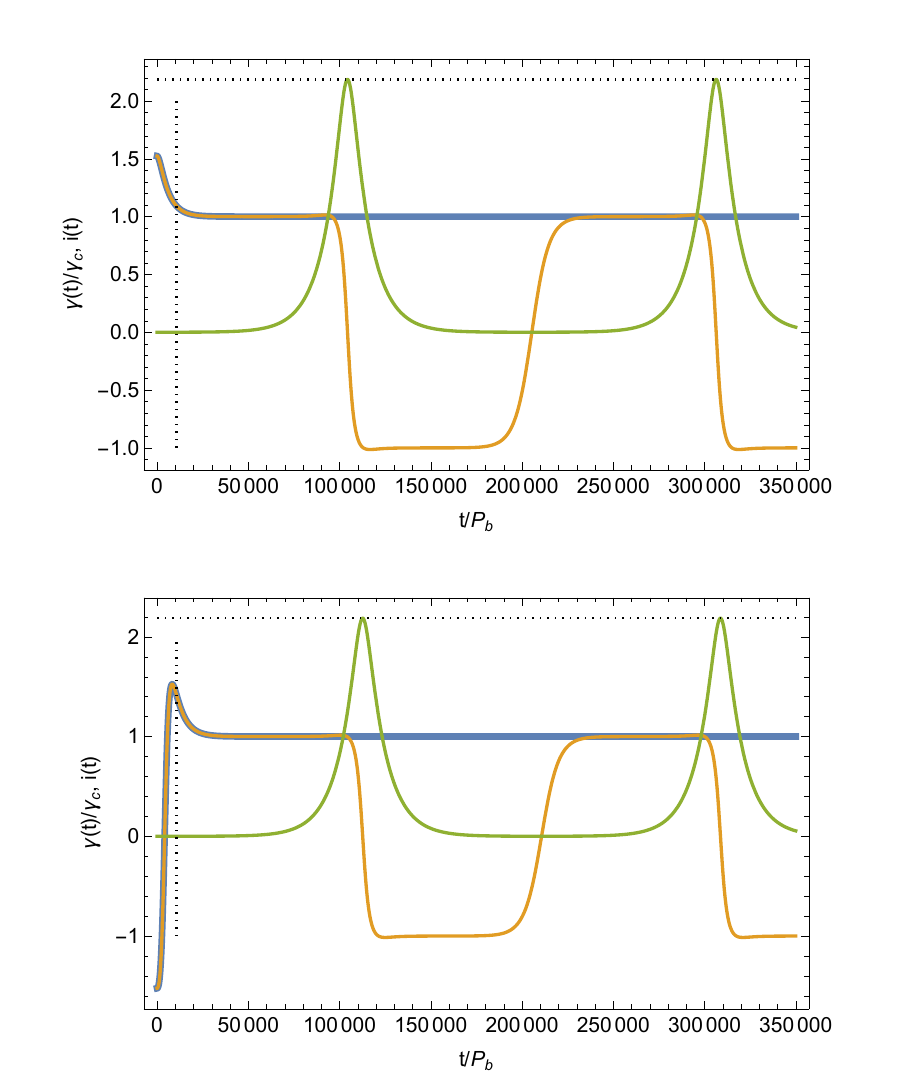}
    \caption{Inclination evolution of an outer planet orbiting  at  $a_4=15\,a_{\rm b}$   around an equal mass binary system of total mass $1 M_{\odot}$ having binary eccentricity $e_{\rm b}=0.8$ and with an inner Jupiter-mass polar planet orbiting at $a_3=5\,a_{\rm b}$.  Time is in units of binary orbital period $P_{\rm b}$. The outer planet begins on a circular orbit and nearly coplanar to the binary orbit with inclination $i_0=0.0001$.
     The upper (lower)  panel is for outer planet initial nodal phase  $\phi_0=45^\circ$ ($\phi_0=-45^\circ$). Plotted in blue is the normalized inclination growth rate determined by the analytic expression for $\gamma(t)/\gamma_{\rm c}$ given by Equation (\ref{gamma}) that is valid until $i_4$ becomes of order unity. Plotted in orange and green are respectively the normalized growth rates $1/\gamma_{\rm c} d \ln{i_4}/dt$ and inclination $i_4$ in radians obtained
    by integrating Equations~(\ref{didt}) and (\ref{dphidt}). The dotted vertical line indicates a time of $1/\gamma_{\rm c}$. The dotted horizontal line is the
    value of maximum inclination given by Equation (\ref{imax}).  The analytically determined curve
    for the growth rate, derived in the small $i_4$ limit,  follows the numerically determined curve until $i_4$ becomes of order unity.
    }
    \label{fig:gri}
\end{figure}

We consider a fiducial model in which $m_1=m_2 =0.5 m_{\rm b}, e_{\rm b}=0.8, m_3= 0.001 m_{\rm b},  a_3=5 a_{\rm b}, a=15  a_{\rm b}$,  $i_0=0.001$, and $e_3=e_4=0$. In this case, Equations (\ref{ai}), (\ref{ao}), and (\ref{imax}), imply that
$a_{\rm i}=12.1 a_{\rm b},  
a_{\rm o}=23.4 a_{\rm b},$ and  $i_{\rm max}= 4.37=125^{\circ}$.
The upper panel of Figure~\ref{fig:gri} plots in blue $\gamma(t)/\gamma_{\rm c}$ given by Equation (\ref{gamma}) for a case with $\phi_0= 45^\circ$. The growth rate initially varies, but settles
to a constant value after a time of about $1/\gamma_{\rm c}$. Also plotted are the results for $\gamma(t)$ (orange) and $i_4(t)$ (green) based on numerical integration of Equations (\ref{didt}) and (\ref{dphidt}) for the same parameters.
As expected, there is nearly exact agreement with the analytic results until $i_4$ in radians is of order unity.

However, for $90^\circ <\phi_0 < 180^\circ$  and for $270^\circ < \phi_0 < 360^\circ$, 
we have that $\sin{(2 \phi_0)} < 0$ and from Equation (\ref{didt}) the inclination initially decays.
%his property is also found for $\gamma(t)$ in Equation (\ref{gamma}). 
%In that case, from Equations (\ref{u1}) and (\ref{u2}), it follows that $u_1(t)$ approaches $u_2(t)$ on a timescale of $\gamma_{\rm c}^{-1}$.
Following an initial decay, the inclination grows. The time at which the $\gamma$ reaches zero, denoted by $t_1$, is determined by Equation (\ref{gamma})
as
\begin{equation}
\gamma_{\rm c} t_1 = \rm{arctanh}\left (\rm{min}\left ( -\frac{\tan \phi_0}{\tan{\phi_{\rm c}}},  -\frac{\tan \phi_{\rm c}}{\tan{\phi_0} }\right) \right). \label{tana}
\end{equation}
Since $\tan \phi_0<0$, we see that $t_1>0$. Notice that for $\phi_0=-\phi_{\rm c}$ or $\phi_0=-\phi_{\rm c}+180^\circ$, time $t_1$ is infinite which is a consequence of the inclination $i_4(t)$ decaying exponentially  at all times at rate $\gamma_{\rm c}$.  The initial time required for the inclination to grow, $t_1$, decreases  as $\phi_0$ departs from $-\phi_{\rm c}$.

The lower panel of Figure~\ref{fig:gri} plots in blue $\gamma(t)/\gamma_{\rm c}$ given by Equation (\ref{gamma}) for a case with $\phi_0= -45^\circ$. The growth rate is initially negative, but settles
to a constant value after a somewhat longer time than $1/\gamma_{\rm c}$. 
As in the case of the upper panel, there is nearly exact agreement between with the analytic results until $i_4$ is of order unity. 
The results in the upper and lower panels are nearly identical, apart from a time shift, after the initial adjustment phase. 

\begin{figure}
	\includegraphics[width=\columnwidth]{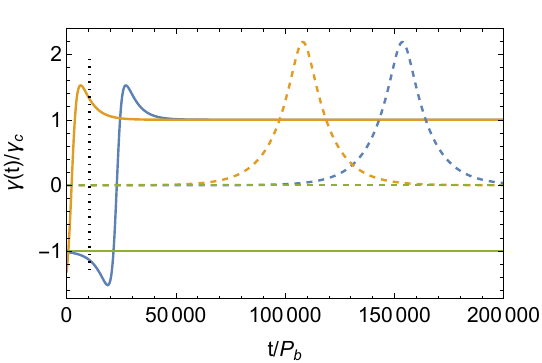}
    \caption{Normalized growths rates  $\gamma(t)/\gamma_{\rm c}$ (solid lines) given by Equation~(\ref{gamma}) and inclination $i_4$ in radians (dashed lines) as a function of time  determined by numerical integration of Equations (\ref{didt}) and (\ref{dphidt})  for the parameters used in Figure~\ref{fig:gri}, but with outer planet initial nodal phase $\phi_0=-\phi_{\rm c}=-0.358=-20.51^{\circ}$ (green), $\phi_0=-0.366 =-21^{\circ}$ (blue), and  $\phi_0= -\pi/3=-60^{\circ}$  (orange),
    The dotted vertical line indicates a time of $1/\gamma_{\rm c}$.}
    \label{fig:grn}
\end{figure}

Figure~\ref{fig:grn} plots other cases that have $-90^\circ < \phi_4 <0$ and initially negative growth rates.
From this figure we see that for $\phi_0=-\phi_{\rm c}$, the inclination has a constant negative growth rate and $i_4(t)$ remains
small at all times plotted, as expected by Equation (\ref{tana}). But for a slightly different value of $\phi_0$, the growth rate evolves to a positive value after
a timescale considerably longer that $1/\gamma_{\rm c}$. The initial peak of inclination is then delayed considerably.
For a more negative initial nodal phase, the growth is less delayed, but occurs in a somewhat longer time than $1/\gamma_{\rm c}$,
as we also found in the lower panel of Figure ~\ref{fig:gri}.
Overall, the delay time to exponential growth is of order $1/\gamma_{\rm c}$
that is a single e-folding growth time during the exponential growth,
%that is of order the apsidal precession timescale $\sim 1/d\varpi/dt$.

%Consequently, we see rthat $\phi_4(t)$ generally approaches $\phi_{\rm c}$ and that $\gamma(t)$ approaches $\gamma_{\rm c}$ on this same timescale  $\gamma_{\rm c}^{-1}$.

 %\begin{figure}
%	\includegraphics[width=\columnwidth]{griec.pdf}
%    \caption{Same as the upper panel Figure~\ref{fig:gri} but with outer planet eccentricity $e=0.5$.
%    }
%    \label{fig:griec}
%\end{figure}

\section{Geometric Interpretation}
\label{sec:geom}

\begin{figure}
	\includegraphics[width=\columnwidth]{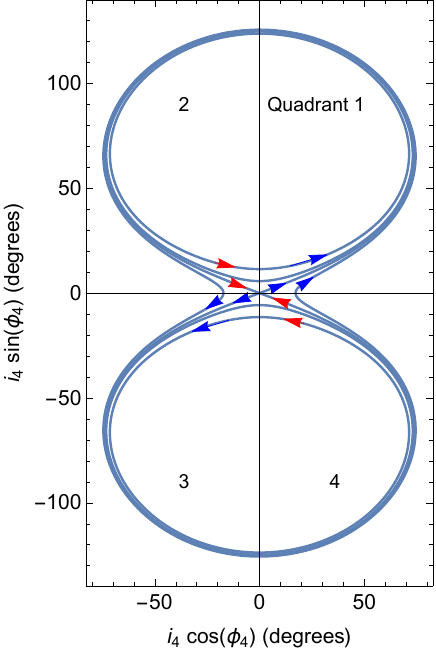}
    \caption{ Phase portrait of   $i_4 \cos{(\phi_4)}$ versus $i_4 \sin{(\phi_4)}$ for the system parameters described in Figure~\ref{fig:gri} based on the numerical integration of Equations (\ref{didt}) and (\ref{dphidt}). A few trajectories are plotted that pass near or through $i_4=0$.
    The blue (red) arrows indicate increasing (decreasing) inclination in time.
     }
    \label{fig:phpl}
\end{figure}

The inclination instability of the outer planet can be understood in terms of a phase portrait as shown in Figure~\ref{fig:phpl}.
The phase portrait plots  $i_4 \sin{(\phi_4)}$ versus $i_4 \cos{(\phi_4)}$ for a few different
orbits that pass close to $i_4=0$. For any point in the plot, 
the value of the outer planet inclination $i_4$ is its distance from the origin and the 
nodal phase $\phi_4$ is its polar angle from the horizontal.  The  binary parameters are  $m_1=m_2=0.5 m_{\rm b}$ and 
 $e_{\rm b}=0.8$.
As the orbit approaches the origin $i_4=0$, as occurs
in quadrants 2 and 4,
its inclination deceases in time, in accordance with Equation (\ref{didt}). This decrease is indicated by red
arrows in the figure. Similarly, as an orbit moves away from the origin as occurs
in quadrants 2 and 4,
its inclination increases in time, in accordance with Equation (\ref{didt}). This increase is indicated by the blue
arrows in the figure. 
The phase trajectory that passes through the origin follows straight lines near the origin with a cusp at $i_4=0$.
These lines lie on the separatrix between the librating orbits that undergo phase variations that are less than $360^\circ$
to circulating orbits that undergo phase variations of $360^\circ$.
In quadrant 1 this line is at angle $\phi_4=\phi_{\rm c}$ given by Equation (\ref{phic}). The inclination increases exponentially
in time along this line at rate  $\gamma_{\rm c}$ given by Equation (\ref{gammac}). The orbit requires infinite time
to depart from the origin along this line. If the orbit starts slightly away from the origin, the inclination grows in finite time
to order unity values. Similarly, the orbit following the straight line in quadrant 4 undergoes exponentially
decreasing inclination and requires infinite time to reach the origin. Orbits off the straight lines never reach $i_4=0$ and always achieve order unity inclination (in radians) in finite time.
The upper panel in Figure~\ref{fig:gri} shows  initially growing inclination $\gamma >0$, since it describes an orbit  that starts
in quadrant 1.
The  lower panels in Figure~\ref{fig:gri} and Figure~\ref{fig:grn} show initially decreasing inclination, since these orbits
starts in quadrant 4. But they later evolve to quadrant 3 where the inclination grows. An exception is the case
of the green line in  Figure~\ref{fig:grn} that  lies along the straight line in quadrant 4 and has only decreasing inclination.
It has phase $-\phi_{\rm c}$ near the origin.

Notice that the orbits that start near the origin but not along the straight lines converge towards the straight
line orbits at later time in quadrants 1 and 3 as the inclination grows.This is consistent with the behavior of the
growth rates in Figures~\ref{fig:gri} and \ref{fig:grn} that show evolution to the exponential growth rate $\gamma_{\rm c}$ for the straight line orbits.

\section{Comparison with numerical simulations}
%Compare  a few simulations with Equations(\ref{imax}), (\ref{gamma}), and possibly (\ref{phit}).
\label{sec:num}

\subsection{Models}
\label{sec:models}

\begin{table}
\begin{center}
\begin{tabular}{ |c|c|c| c| c| c|  c|  c|} 
\hline
Model & $e_{\rm b}$  & $e_4$ & $q_{\rm b}$ & $a_{\rm i}/a_{\rm b}$ & $a_{\rm o}/a_{\rm b}$ & $i_3(0)$  (deg) & $i_0 $(deg) \\
\hline
\hline
A1 & 0.2 & 0 &  1 & 14.0 &  14.8 &  88.7 &0 \\ 
A2 & 0.2 & 0.5 & 1 & 16.5 &  17.4 &  88.7  &0\\
A3 & 0.2 & 0.5 &0.5 & 15.9 & 16.8 & 88.7&0\\
B1 & 0.5 & 0 &  1 & 13.5 &  17.8   &  89.3 &0 \\ 
B2 & 0.5 & 0.5 & 1& 15.9 &  21.0 &   89.3 &0 \\
B3 & 0.5 & 0.5 &0.5  & 15.4 & 20.3  &  89.3 &0 \\
C1 & 0.8 & 0 &  1 &12.1 & 23.4 & 89.7 &0\\ 
C2 & 0.8 & 0.5 & 1 &14.3  & 27.6 &  89.7  &0  \\
C3 & 0.8 & 0.5 &0.5 &13.8 &  26.6  &  89.7 &0\\
C4 & 0.8 & 0 &  1 &12.1 & 23.4 & 89.7 & 2\\ 
D1 & 0.8 & 0 & 0.5 & 11.7 & 22.6 &  89.7 &0 \\
D2 & 0.8 & 0 & 0.1 & 8.8 & 17.0 & 89.2 &0 \\
D3 & 0.8 & 0 & 0.01 & 4.8 & 9.3 &  83.4 &0 \\
E1 & 0.8 & 0 & 0.1 & 8.8 & 17.0 & 89.2 & 1 \\
E2 & 0.8 & 0 & 0.01 & 4.8 & 9.3 &  83.4& 1 \\
\hline
\end{tabular}
\end{center}
 \caption{Simulation parameters. The first column is the simulation name. The second column is the binary eccentricity. The third column is the eccentricity of the outer planet. The fourth column is the binary mass ratio. The fifth column is the predicted inner semi-major axis
 of the unstable region given by Equation (\ref{ai}) and the sixth column is the predicted outer semi-major axis
 of the unstable region given by Equation (\ref{ao}).  The seventh column $i_3$ is the initial inclination of the inner planet relative
 to the orbital plane of the binary in the generalised polar state.  The eighth column $i_0$ is the initial inclination of the outer planet relative
 to the orbital plane of the binary. 
}
\label{table}
\end{table}

We compare our analytic model with numerical simulations using the $n$-body code {\sc rebound} \citep{Rein2012}. We apply a set of models with parameters given  by Table~\ref{table}.  The remaining  parameters are described below.
For all these models the inner planet has a mass of $0.001 m_{\rm b}$,   an initial semi-major axis of $a_3=5 \, a_{\rm b}$, and is in generalized polar orientation \citep{Martin2019}.
In a generalized polar orientation, the orbit of the outer planet is stationary in a frame that precesses with the binary. 
The longitude of the ascending  node is equal to $90^\circ$, relative to the eccentricity vector of the binary.
 We determine
the inclination for the generalized polar orientation to quadrupole order in the binary potential. A zero mass planet has a stationary orbit that is polar, with a tilt that is perpendicular to the binary orbital plane \citep[e.g.,][]{Farago2010}.  But for a planet with nonzero mass, as we have here, this stationary
orbit %in the frame that precesses with the binary 
has a smaller tilt.  Due to the nonzero inner plane mass, the binary has
a small tilt relative to the invariable plane (the plane perpendicular to the total angular momentum of the system). In {\sc rebound}, we apply a reference direction that is in the invariable plane and along the initial eccentricity vector of the binary projected onto that plane. For all models, the initial
longitude of the ascending node for the binary and outer planet is $-90^\circ$ and for the inner planet is $90^\circ$. The initial argument of periapsis for all eccentric
orbits is $90^\circ$.
Unless otherwise stated, we present all inclinations as relative to the instantaneous orbital plane of the binary.
 We employ the IAS15 integrator \citep{Rein2012}.

\subsection{Inclination Oscillations}
\label{sec:inclosc}
 
\begin{figure}
	\includegraphics[width=\columnwidth]{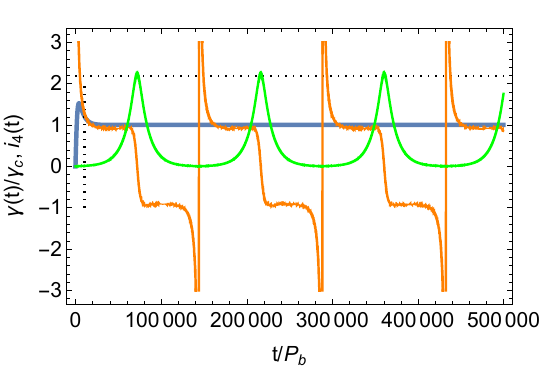}
    	\includegraphics[width=\columnwidth]{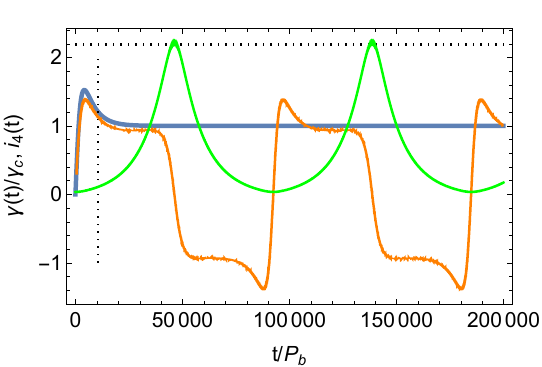}
    \caption{Plot of the inclination evolution of a massless outer  planet with initial semi-major axis $a=15 a_{\rm b}$ for Model C1 (upper panel) and C4 (lower panel).
    Plotted as green solid lines  are inclination values $i_4$ in radians obtained
    from {\sc rebound} simulations.  Plotted as the orange solid lines  are
    the normalized growth rates $1/\gamma_{\rm c} d \ln{i_4}/dt$ using $i_4$
    from {\sc rebound} simulations.
    The blue solid lines are the normalized inclination growth rates determined by the analytic expression for $\gamma(t)/\gamma_{\rm c}$ given by Equation (\ref{gamma}) that is valid until $i_4$ in radians becomes of order unity.  The dotted vertical line indicates a time of $1/\gamma_{\rm c}$. The dotted horizontal  line is the predicted
    value of maximum inclination in radians given by Equation (\ref{imax}). }
    \label{fig:grin}
\end{figure}

We consider the Models C1 and C4 in Table~\ref{table} 
with an initial semi-major axis of the outer planet of 
$15 a_{\rm b}$. This value of $a$ lies between $a_{\rm i}$ and $a_{\rm o}$ and
therefore $i_4$ is then expected to grow.
The two models have the same parameters but differ
in that for Model C1 the initial orbit of the outer planet is coplanar with the binary orbital plane (to numerical accuracy), while for Model C4 it is inclined by $2^\circ$.
As expected, the numerical simulations show that the orbit of the outer planet undergoes inclination oscillations as plotted by green lines
in Figure~\ref{fig:grin}.  The predicted maximum value (dotted horizontal line) agrees well with the peak values of the green lines in the simulations.

Plotted as blue lines in Figure~\ref{fig:grin} are the initial time dependent growth rates predicted by Equation (\ref{gamma}) with initial longitude 
of ascending node of $\phi_0=-90^\circ$. Recall that these rates are valid until $i_4$  in radians becomes of order unity. Plotted as orange lines are the
inclination growth rates obtained from the simulations.
 However, for Model C1 (top panel) at early times $t < 1/\gamma_{\rm c}$, there is disagreement between the analytic and
 numerical growth rates. 
The analytic model assumes that
the binary lies in the invariable plane. But due to the small mass outer planet
there is a small tilt $\sim 0.8^\circ$ of the binary orbit away from this plane. 
We attribute
this disagreement with that difference. For larger tilts of the outer planet
the difference between the planes becomes less important to the inclination evolution. 
At later times the growth rates agree well over the inclination growing phase while $i_4 \la 1$.
For Model C4 that has a larger initial inclination
there is good agreement in the growth rate evolution at all early times.

\subsection{Effect of binary eccentricity}
\label{sec:eb}

\begin{figure}
	\includegraphics[width=\columnwidth]{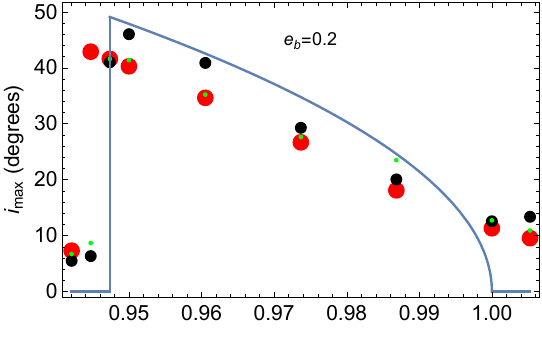}
	\includegraphics[width=\columnwidth]{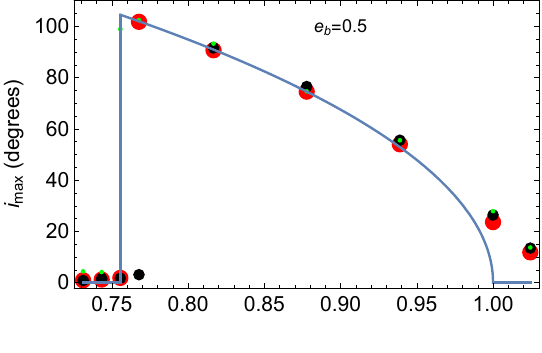}
	\includegraphics[width=\columnwidth]{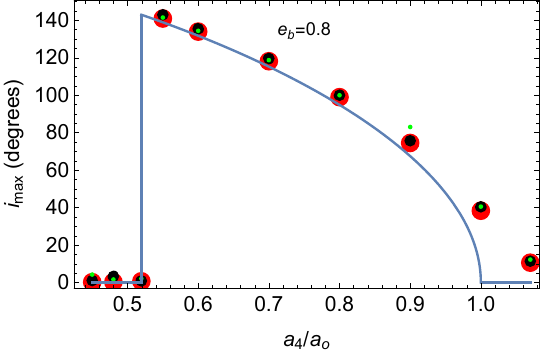}
    \caption{Maximum inclination of the outer planet relative to the binary orbital plane as a function of outer planet initial  semi-major axis normalized by
    $a_{\rm o}$, the predicted  outer semi-major axis of the  unstable region, for three values of initial binary eccentricity.  The {\sc rebound} simulation results are plotted by large red circles for  
 Models A1, B1, and C1, by black circles for Models A2, B2, and C2, and small green circles for Models A3, B3, and C3. The blue sold lines plot Equation (\ref{imax}) of the analytic model.}
    \label{fig:imax}
\end{figure}

In Figure~\ref{fig:imax} we plot the maximum inclination achieved for the outer planet for Models An, Bn, and Cn for $n=1,2,3$ listed in Table~\ref{table} as a function of the initial semi-major axis of the outer planet $a_4$ normalized by $a_{\rm o}$. For these cases the orbit of the outer planet is initially coplanar with the orbit of the binary. Three different initial values of binary eccentricity are considered.
The values of $a_4/a_{\rm o}$  on the horizontal axis can be easily transformed to $a_4/a_{\rm b}$ 
by applying the scaling
factor $a_{\rm o}/a_{\rm b}$ that is provided for each model in Table~\ref{table}. For example, for the red dots in the bottom panel Figure~\ref{fig:imax} (Model C1), we have from Table 1 that the horizontal axis is multiplied by 23.4 to obtain $a_4/a_b$. 
The  simulations are run for $10^6 P_{\rm b}$, where $P_{\rm b}$ is the binary orbital period.
As noted in Section~\ref{sec:ana},  in  the case that  the outer planet is on a circular orbit or the binary has equal mass members,  the octupole order terms vanish in  the Hamiltonian and the analytic model should be more accurate.
Models A1, B1, and C1 have $e_4=0$ and $q_{\rm b}=1$ and a vanishing octupole term.
Models A2, B2, and C2 have $e_4=0.5$ and $q_{\rm b}=1$   and a vanishing octupole term.
Models A3, B3, and C3 have $e_4=0.5$ and $q_{\rm b}=0.5$  and a nonvanishing octupole term.
For given binary eccentricity, we ran simulations with a set of values for $a_4/a_{\rm o}$ that is fixed for the three models,
$n=1,2,3$. 
 The analytic model given by Equation (\ref{imax}) plotted by the  solid blue lines predicts that the maximum value of the inclination
is independent of $n$ for a given model type (A, B, and C), for fixed $x=a_4/a_{\rm o}$.  
Therefore, the three markers plotted by the  red circle ($n=1$),  black circle ($n=2$),
and green circle ($n=3$) should overlap. 

Figure~\ref{fig:imax}  shows  overall good agreement between the analytic model and the simulations.
The agreement is less good at low binary eccentricity $e_{\rm b}=0.2$ where the range of unstable radii
is small.  The agreement is best at high binary eccentricity  $e_{\rm b}=0.8$, where the range of unstable radii
is large. In that case, the circles overlap well and closely track the transition in $a_4/a_{\rm o}$ from stable ($i_{\rm max}=0$) to unstable inclination at $a_4 = a_{\rm i}$ . 
The case  with intermediate binary eccentricity  $e_{\rm b}=0.5$ shows good agreement but less so than the case 
 with $e_{\rm b}=0.8$. %\RGM{I could go either way on this, but I wonder if it would be easier to compare the panels if the scales were the same?}
 
\begin{figure}
	\includegraphics[width=\columnwidth]{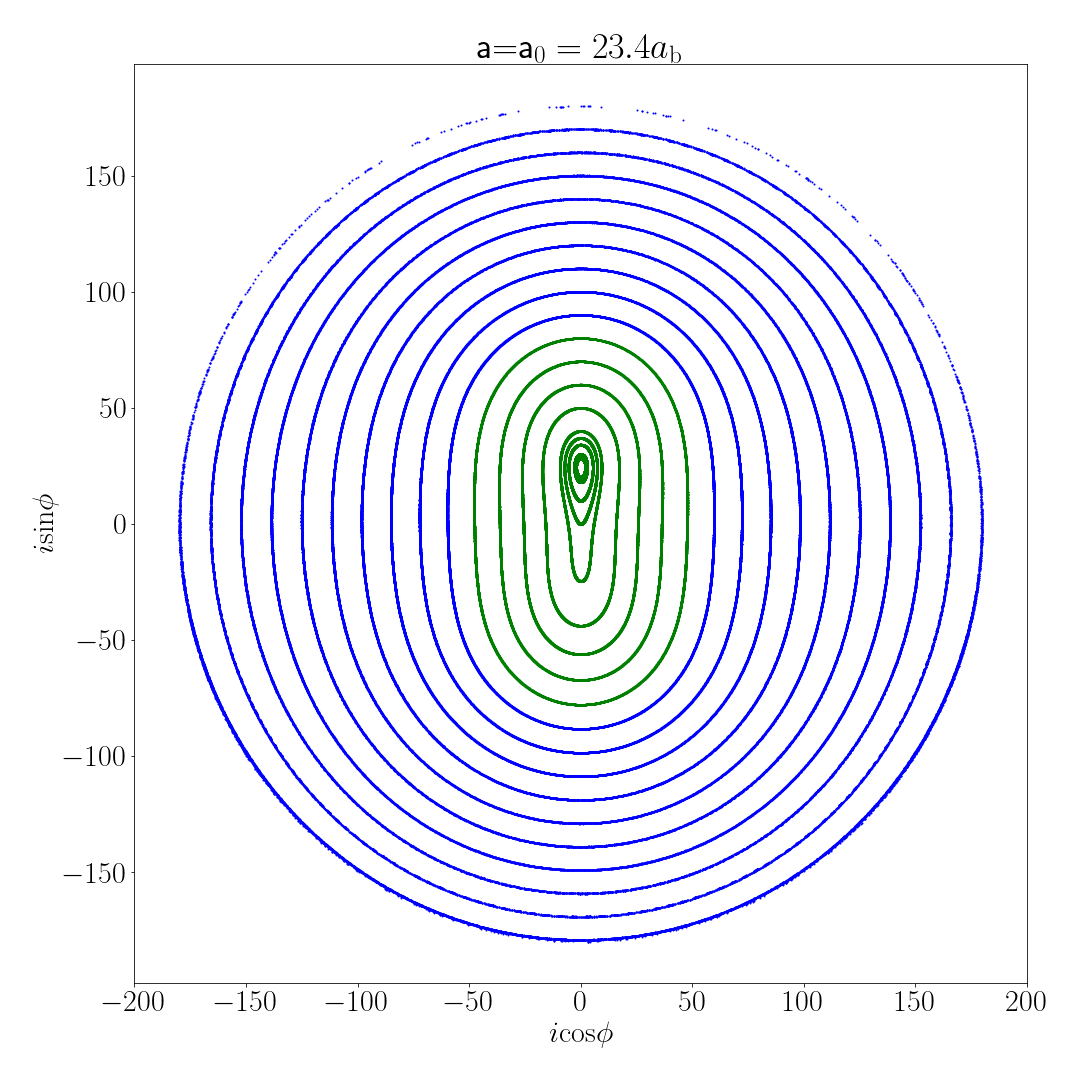}
    \caption{ Phase portrait in degrees for the outer planet in a system  with Model C1 parameters and outer planet orbital semi-major axis  at $a_4= a_{\rm o}= 23.4 a_{\rm b}$, where the analytic model predicts that
    the maximum inclination of an initially nearly coplanar orbit remains close to zero.}
    \label{fig:phaseao}
\end{figure}

For all models at $a_4 \ge a_{\rm o}$ the plotted values of maximum inclination are nonzero, while
the analytic model predicts them to be zero.  In the case of $e_{\rm b}=0.8$, 
we  have  from the simulations that $i_{\rm  max} = 37^\circ$ at   $a_4=a_{\rm o}$  which drops to 
$i_{\rm  max} = 10^\circ$ at $a_4=1.07 a_{\rm o}$. We have investigated this effect and find that the orbits
for $a_4 \ga a_{\rm o}$  are quite different from those described by the analytic model.
Figure \ref{fig:phaseao} shows the phase portrait of orbits at the outer  semi-major axis of the unstable region 
$a_{\rm o}$.
The plot shows that there is an asymmetry in that the orbits are not symmetric between the upper
and lower half planes, such as as shown in Figure~\ref{fig:phpl}. 
The transition from resonant to nonresonant orbits near $a_4=a_{\rm o}$
involves a shift of the libration centers in both the upper and lower half planes to $i_4=0$.  
The plot shows a libration center occurs in the upper half-plane, but not in the lower half plane. This asymmetry is reversed if the initial nodal phase of the inner planet is changed by $180^\circ$.  We attribute this effect to nonzero angular momentum of the
inner planet that causes a small change in tilt of the binary orbit that is not taken into account
in the analytic model. %Although this change in tilt is small, $i_{\rm max}$ varies rapidly
%with $a$ near $a= a_{\rm o}$. 
%The libration centers of the orbits in the upper and lower half planes may not reach $i=0$ at the same orbital radius $a_{\rm o}$ that is predicted by the analytic model.

\subsection{Effect of the initial semi-major axis of the outer planet}
\label{sec:sma}

\begin{figure}
\includegraphics[width=\columnwidth]{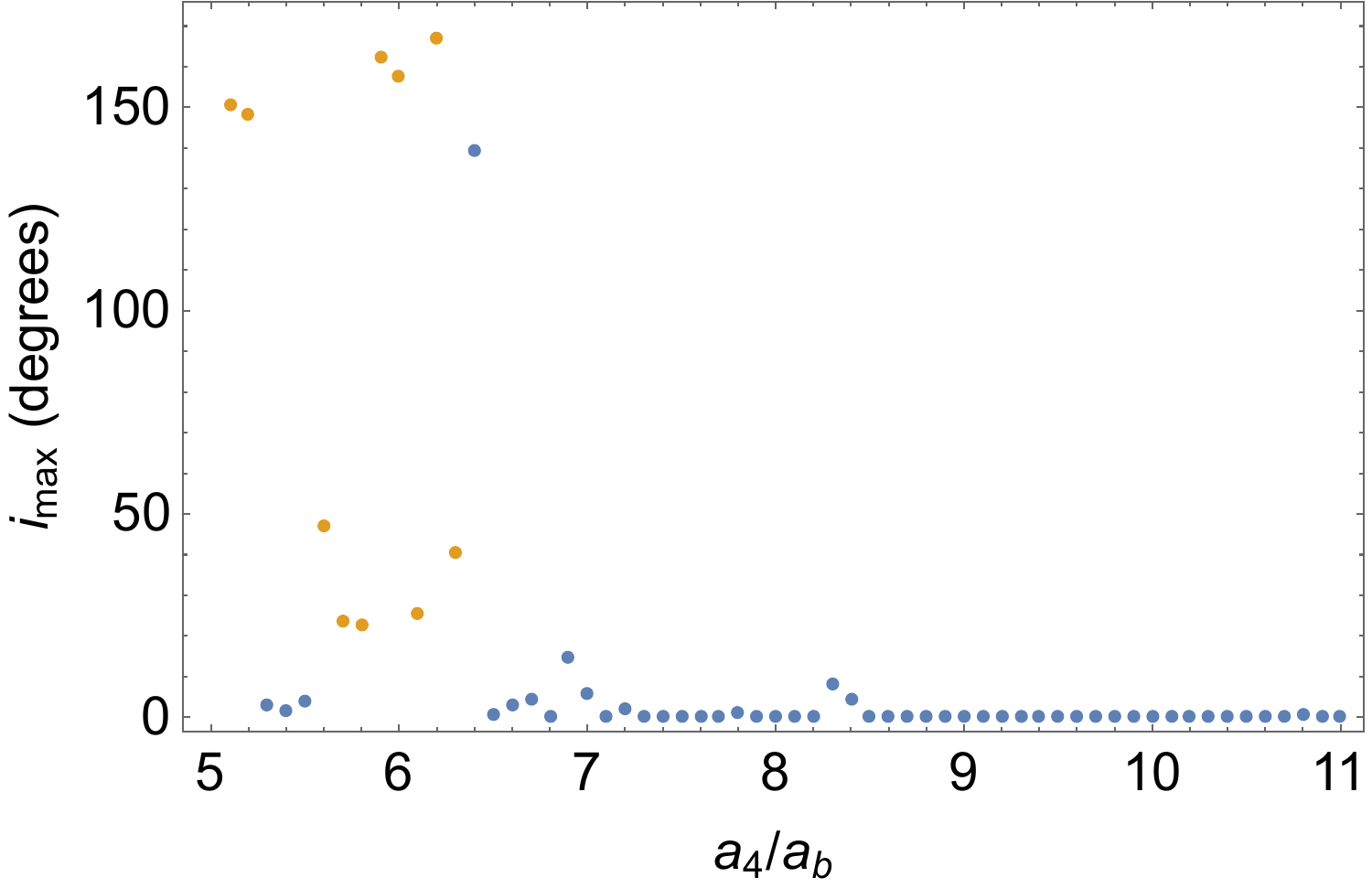}
    \caption{Plot of the outer planet maximum inclination relative to the binary orbital plane  determined by {\sc rebound} simulations for parameters in Model C1 as a function of initial semi-major axis normalized by $a_{\rm b}$.  The gold dots are for cases where the outer planet becomes unbound    from the system, while the blue dots are for bound cases.  }
    \label{fig:imaxnby}
\end{figure}

As noted in Section \ref{sec:ana}, we have assumed that the two planets are well enough separated  so that
the outer planet's evolution is controlled by its interaction with the binary and that its interaction with the inner
planet can be ignored. The agreement in Figure~\ref{fig:imax} between the analytic and numerical values
for the maximum inclination indicates that this assumption holds for $a_4$ values in which the inclination instability occurs, $a_{\rm i} < a_4 < a_{\rm o}$,
in the models we considered. 
%  \cite{Childs2023} give a criterion for the  analytic model breaking down when 
 %the nodal precession rate of the outer planet due to the inner planet is greater than
 %the nodal precession rate due to the binary. 
 The orbits of the two planets are generally in different planes.
However, if $a_3 \simeq a_4$,  the two planets can interact strongly and the model in Section  \ref{sec:ana}
can break down.  

We explore this breakdown for Model C1 by extending the results plotted as red circles in the bottom 
panel of Figure~\ref{fig:imax} to smaller initial values of $a_4$. Since such cases involve $a_4 < a_{\rm i}$,
the analytic model predicts that $i_{\rm max}=0$.  Figure \ref{fig:imaxnby} plots the values of $i_{\rm max}$ 
as a function of the outer planet's initial semi-major axis based on simulations.
Gold dots denote cases where the outer planet becomes unbound from the system and blue dots denote cases where the outer planet remains bound over $10^6 P_{\rm b}$. As seen from the plot, the breakdown occurs for $a_4 \la 7.0 a_{\rm b}$,
where the outer planet is frequently ejected from the system. At larger values of $a_4$, the inclination remains close to its
initial value of zero. There are some deviations at $a_4 \simeq 7 a_{\rm b}$ and $a_4 \simeq 8.3 a_{\rm b}$.
The latter case may involve the 2:1 resonance between the planets that occurs at $a_4 \simeq 8.1 a_{\rm b}$.
For the analytic model to hold, we require $a_{4 \rm{cr} } < a_{\rm i}$,  where $a_{4 \rm{cr}}$ is the critical semi-major axis of the outer planet  for breakdown.
From Equation (\ref{ai}), this requirement is then
\begin{equation}
 \sqrt{ \frac{a_{\rm b}} { a_3} }
 \frac{ \sqrt{1-e_{\rm b}^2 }}{ (1-e_4^2)^2}\, \frac{m_1 m_2}{m_{\rm b} m_3}  > 3 \chi^{7/2},
\end{equation}
where $\chi = a_{4 \rm{cr}}/ a_3$.  In the case of Figure~\ref{fig:imaxnby}, we have $\chi \simeq 7/5 = 1.4$ and this inequality is well satisfied,
as expected. More generally, this inequality 
should be satisfied if the object with  mass $m_3$ is of planetary mass and is orbiting a  binary star system with order unity mass ratio, provided that $e_{\rm b}$ is not very close to unity and that the inner planet is not very far from the binary. %Under such conditions, the analytic model should generally hold well.
Some effects that occur outside this parameter range for extreme binary mass ratios are explored in the next subsection.

\subsection{Effect of binary mass ratio}
\label{sec:q}
 The approximations we have made
in deriving the analytic model break down for sufficiently small binary mass ratio. 
The model assumes that  the gravitational effects of the inner planet on the outer planet
are much smaller than the gravitational effects of the binary.
 \cite{Childs2023} give a criterion for the  analytic model breaking down when 
 the nodal precession rate of the outer planet due to the inner planet is greater than
 the nodal precession rate due to the binary. This criterion can be used 
 to provide a limit on the binary mass ratio
\begin{equation}
q _{\rm b} > q_{\rm prec},
\end{equation}
where for $q_{\rm prec} \ll 1$
we have that 
\begin{equation}
q_{\rm prec}  =  \left( \frac{a_3}{a_{\rm b}} \right)^2 \frac{m_3}{m_{\rm b}} \frac{ \tan{i_4}}{1 + 3/2 \, e_{\rm b}^2}.
\end{equation}

We apply a set of models with $e_{\rm b}=0.8$ and different binary mass ratios 
The models we consider are Models C1, D1, D2, D3, E1, and E2 listed in Table~\ref{table}. 
For all these models with $\tan{i_4} \sim 1$, we have that  $q_{\rm prec} \sim 0.01$.

The analytic model 
ignores the interaction between the two planets. The outer semi-major axis for instability $a_{\rm o}$ decreases
for small mass ratio but must be larger than the inner planet  semi-major axis $a_3$  for the  model to apply. 
For  $a_4 \simeq a_{\rm o}  \simeq   a_3$, strong interactions are expected to
occur between the planets which are initially on  mutually orthogonal orbits and the tilt instability described here would not apply. For  $a_{\rm b} \ll a_4 \ll a_{\rm o}  \simeq a_3$,
the planet could instead be subject to Kozai-Lidov oscillations \citep{Kozai1962,Lidov1962} but modified by the inner binary \citep[e.g.,][]{Martin2022}.
This requirement,
$a_{\rm o} > a_3$, implies a condition of the binary mass ratio
\begin{equation}
q _{\rm b} > q_{\rm ppo},
\end{equation} 
where from Equation (\ref{ao}),  for $q_{\rm ppo} \ll 1$, we have that
\begin{equation}
q_{\rm ppo} = \frac{3 \left(\frac{m_3}{m_{\rm b}} \right) (1-e_4^2)^2  \sqrt{\frac{a_3}{a_{\rm b}}}  \sqrt{1-e_{\rm b}^2}}{1+ 4 e_{\rm b}^2}.
\label{qmin1}
\end{equation}
%We apply a set of models with $e_{\rm b}=0.8$ and different binary mass ratios 
%for which the outer planet has $a_4=a_{\rm m}$, the orbital semi-major axis  of maximum growth rate given by Equation (\ref{am}). 
For the models we consider, Models C1, D1, D2, D3, E1, and E2, we have that  $q_{\rm ppo} = 0.0011$.

We consider instead the  weaker requirement that the semi-major axis of the outer planet for maximum growth rate 
be greater than the semi-major axis of the inner planet. This requirement,
$a_{\rm m} > a_3$, implies that
\begin{equation}
q_{\rm ppm} = \frac{3 \left(\frac{m_3}{m_{\rm b}} \right) (1-e_4^2)^2  \sqrt{\frac{a_3}{a_{\rm b}}}   (2+ 3 e_{\rm b}^2)}{2 (1+ 4 e_{\rm b}^2)\sqrt{1-e_{\rm b}^2} } \label{qmm} 
\end{equation}
for $q_{\rm ppm} \ll 1$.

\begin{figure}
	\includegraphics[width=\columnwidth]{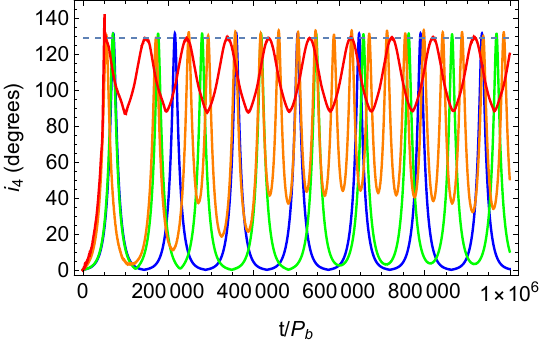}
    \caption{Solid lines plot the outer planet inclination evolution relative to the binary orbital plane determined by {\sc rebound} simulations with different binary mass ratios. The outer planet orbit  is initially coplanar  with the binary orbital plane. The binary mass ratios
    are: $q_{\rm b}=1$  (blue, Model C1),  0.5 (green, Model D1),  0.1 (orange, Model D2), and 0.01 (red, Model D3). For each case, the value of the initial semi-major axis is 
    $a_4 =a_{\rm m} = 0.616 a_{\rm o}$ given by  Equation (\ref{am}).  The horizontal dashed line is the  predicted maximum inclination given by Equation (\ref{imax})  that has the same value for all these cases. }
    \label{fig:q}
\end{figure}

We carried out simulations with $a_4=a_{\rm m}$ for Models C1, D1, D2, D3, E1, and E2.
For all these models, we have that  $q_{\rm ppm} = 0.006$.
and Equation (\ref{imax}) predicts that the maximum tilt is $i_{\rm max}= 129.2^\circ$.
Figure~\ref{fig:q} plots the inclination evolution. for a set of models (Models C1, D1, D2, and D3)  that begin with an outer planet  orbit that is coplanar with the binary orbital plane. 
 For all these models, the maximum inclination agrees well with the predictions of the analytic model. For $q_{\rm b}=1$ and $q_{\rm b}=0.5$ the inclination oscillations have minima  that
are close to a tilt of zero. 

For $q_{\rm b}=0.01$ and $q_{\rm b}=0.1$ the inclination minima are not close to zero. 
%For both of these models, the binary mass ratio is of order or smaller than
%the minimum values estimated above based on the precession criterion, $q_{\rm b} %\la  q_{\rm prec} \sim 0.06$.
For $q_{\rm b}=0.01$, the binary mass ratio is of order of
the minimum values  based on the precession and planet-planet interaction criterion, $q_{\rm b} \la  q_{\rm prec} \sim q_{\rm ppm}$. This is expected to lead to dramatically different changes in the orbital properties of the outer planet than the cases of higher values of $q_{\rm b}$, as we show
later.

\begin{figure}
	\includegraphics[width=\columnwidth]{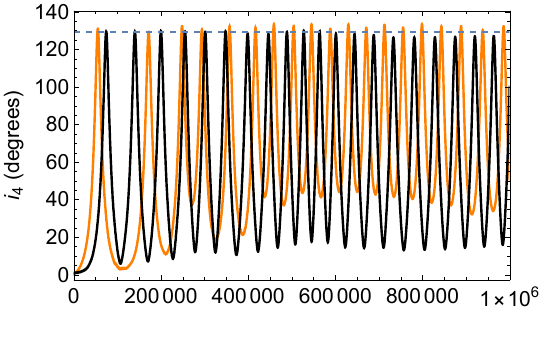}
	\includegraphics[width=\columnwidth]{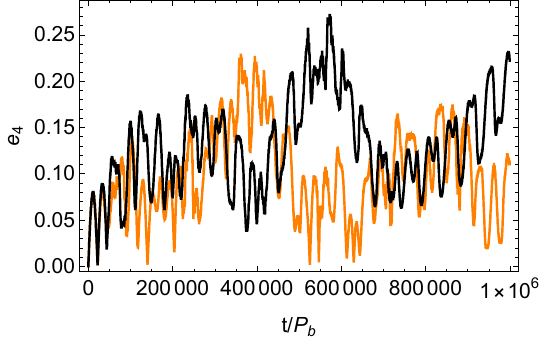}
    \caption{The upper panel plots the outer planet inclination evolution relative to the binary orbital plane  determined by {\sc rebound} simulations with binary mass ratio $q_{\rm b}=0.1$ for two initial inclinations.  The lower panel plots the eccentricity evolution. 
    The orange line is for an  outer planet in Model D2 whose orbit is initially coplanar with the binary orbital plane. The solid black line is for an  outer planet in Model E1 whose orbit is initially tilted by $1^\circ$ with respect to the binary orbital plane. The horizontal dotted line in the upper panel  is the  predicted maximum inclination.  }
    \label{fig:q0p1i}
\end{figure}

Figure~\ref{fig:q0p1i} compares the orbital evolution of two cases with $q_{\rm b} =0.1$ with the same initial parameters but with different initial inclinations. Plotted in orange is the initially coplanar case of Model D2, also plotted in orange in Figure~\ref{fig:q},
while  plotted in black is Model E1 with a small initial inclination of $1^\circ$.
The upper panel shows that the inclination minima are reduced considerably in Model E1. 
Secular theory in quadruple order predicts that the
eccentricity of a circumbinary planet, ignoring planet-planet interactions, is constant that in this case should be zero.
The eccentricity evolution of the outer planet shown in the lower panel
is considerably different in the two cases. The eccentricities undergo complicated variations in time. 
In addition we find that the inner planet also acquires an oscillatory eccentricity with  $0 \le e_3 \la 0.2$.
The eccentricity  $e_3(t)$ is identical in the two models because the outer planet has zero mass.
We also find that the semi-major axis changes are very small $\sim 1\%$ in both cases.
The sensitivity of the inclination minima to the initial inclination of the outer planet suggests that
in this case the system is subject to the effects of a more complicated form of secular instability than given by our analytic model.

The case of Models D3 and E2 with $q_{\rm b}=0.01 \sim q_{\rm ppm}= 0.006$  given by Equation (\ref{qmm}) 
involves strong interactions between the two planets, resulting in a change  in the motion of the outer planet. In these cases, the  initial planet semi-major axis  is $a_4= 1.14 a_3$.  These two models again differ only in initial inclination of the outer planet. 
In Model D3, the outer planet orbit is initially coplanar with respect to the binary orbital plane, while in Model E2 the outer planet is inclined by $1^\circ$. 

\begin{figure}
\includegraphics[width=\columnwidth]{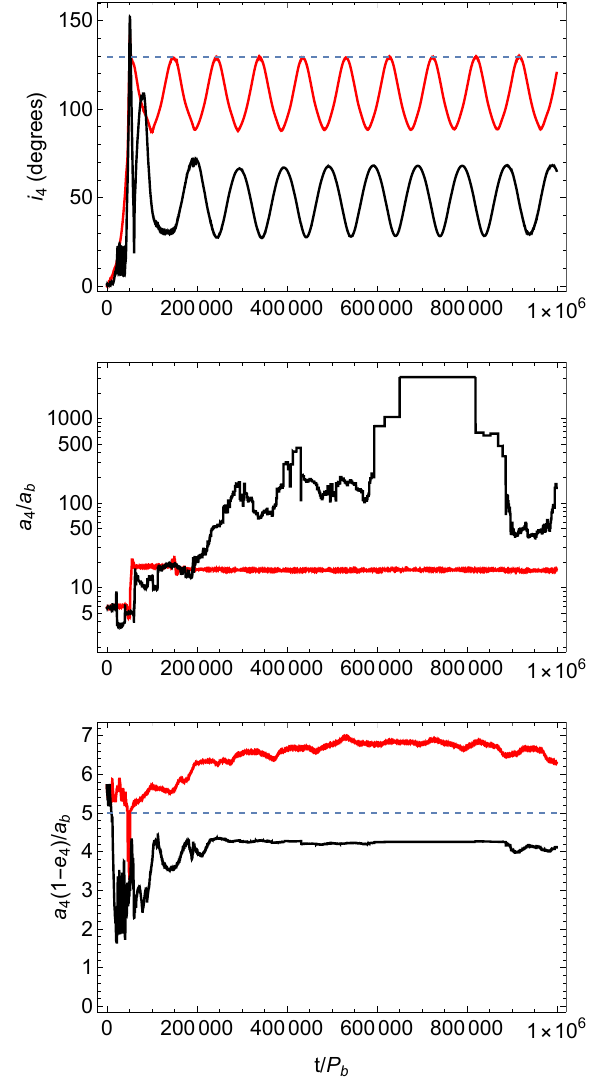}
    \caption{The top panel plots the outer planet inclination evolution relative to the binary orbital plane determined by {\sc rebound} simulations with binary mass ratio $q_{\rm b}=0.01$ for two initial inclinations.  The red line is for an outer planet in  Model D3 whose orbit is initially coplanar with the binary orbital plane.   The solid black line is for an  outer planet  in Model E2 whose orbit is initially tilted by $1^\circ$ with respect to the binary orbital plane.  The middle panel plots the semi-major axis evolution of the outer planet on a logarithmic scale. 
    The bottom panel plots the periastron evolution of the outer planet. 
     The horizontal dashed line in the top panel is the  predicted maximum inclination. The horizontal dashed line
  in the bottom panel is the semi-major axis of the inner planet, $a_3/a_{\rm b}$. }
    \label{fig:q0p01i}
\end{figure}

Figure~\ref{fig:q0p01i} compares the orbital evolution of Models D3 and E2. Plotted in red is the case of Model D3, also plotted in red in Figure~\ref{fig:q},
while  plotted in black is Model E2.
The upper panel of Figure~\ref{fig:q0p01i} shows that inclination oscillations  occur in both cases but with different maxima
and minima.
Therefore, the good agreement between the analytic and simulation results in the inclination maxima for the red line
in Figure~\ref{fig:q} is not a general outcome and is sensitive to initial conditions. 
The middle panel plots the
semi-major axis evolution of the outer planet in the two cases.
The semi-major axis of the inner planet in these cases remains nearly
constant, while its eccentricity undergoes mild oscillations with  $0 \le e_3 \la 0.07$.
 In the initially coplanar case (red), there is a close encounter at a 
time $\sim 5 \times 10^4 P_{\rm b}$ that causes a jump in its semi-major axis that then remains fairly constant.
In the initially inclined case (black), there is an jump in $a_4$ at an early time and
a continual change in $a_4$ to large values. 

As seen in the bottom panel of Figure~\ref{fig:q0p01i}, the periastron of the outer planet in red temporarily drops below 
the semi-major axis value for the inner planet. The orbits lie in different planes, so the orbits are not necessarily close. But in this case the orbital inclinations are nearly equal at the time of equal semi-major axis values.
In the case of the outer planet shown in black, its periastron
soon drops below 
the semi-major axis of the inner planet and remains in that state while its apastron is
well beyond the semi-major axis of the inner planet. This leads to strong interactions  with
the inner planet and binary over long timescales
which results in further energy and semi-major axis changes to the orbit of the outer planet.

\begin{figure}
\includegraphics[width=\columnwidth]{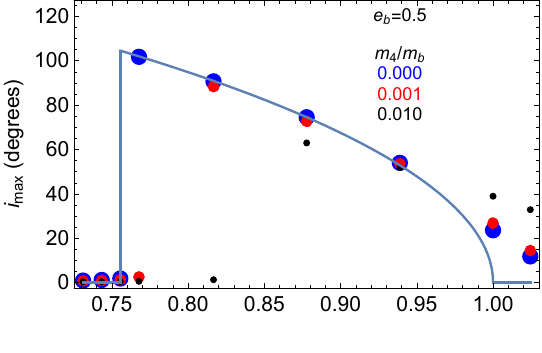}
\includegraphics[width=\columnwidth]{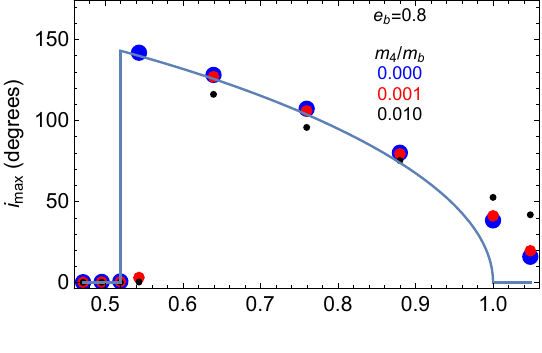}
\includegraphics[width=\columnwidth]{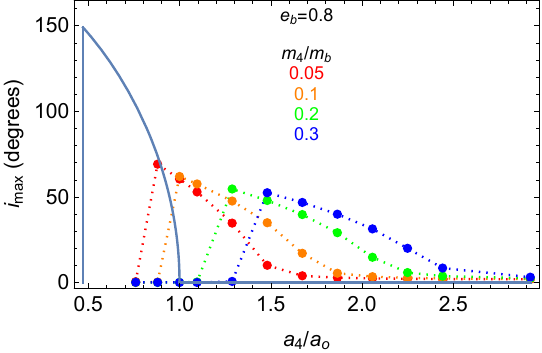}
    \caption{The top and middle panels plot as dots the outer planet maximum inclination relative to the binary orbital plane  determined by {\sc rebound} simulations for parameters in Models B1 and C1, respectively, as a function of initial semi-major axis normalized by $a_{\rm o}$, the predicted outer semi-major axis of the unstable region for a zero mass outer planet.  The bottom panel is the same as the middle panel but for higher masses of the outer object. The dotted lines connect the dots. The blue sold lines plot Equation (\ref{imax}) of the analytic model. The circles in the bottom panel
    sometimes overlap for $i_{\rm max} =0$.
      }
    \label{fig:imaxmeb}
\end{figure}

%\section{Discussion}
%\label{sec:disc}

%\begin{multicols}{1}
\begin{figure*}
%\onecolumn\includegraphics[width=\columnwidth]{HighMass.pdf}
\includegraphics[width=2\columnwidth]{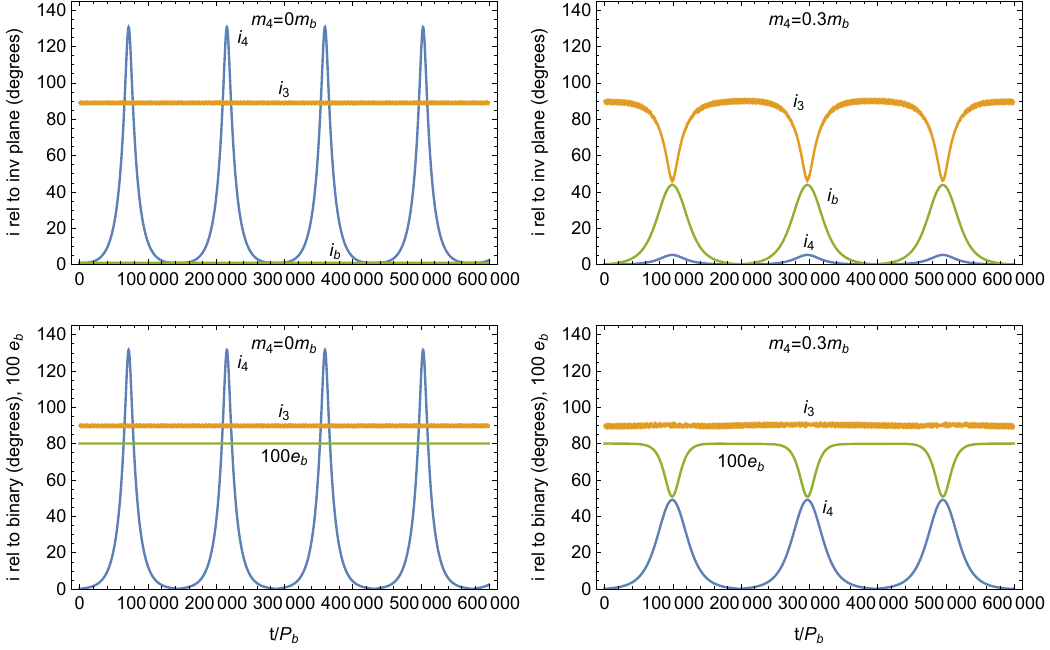}
    \caption{The panels plot the orbital element evolution  determined by 
    {\sc rebound} simulations for parameters of Model C1 with $m_4=0$, $a_4=14.4 a_{\rm b}$ (left) 
    and $m_4=0.3 m_{\rm b}$, $a_4=37.4 a_{\rm b}$  (right).  The upper panels plot the inclination evolution relative to the invariable plane.
    The lower panels plot the inclination evolution relative to the binary orbital plane and also plots the binary eccentricity 
    evolution.}
    \label{fig:HighMass}
\end{figure*}
%\end{multicols}{1}

\subsection{Effect of the outer object mass}
\label{sec:mo}

Up to this point we have considered the outer planet to be a test particle, $m_4=0$.
In this section we analyze the effects of having $m_4 >0$. 
We consider models with parameters given by Model B1 and C1 of Table~\ref{table} and outer object
masses of $m_4/m_{\rm b} = 0, 0.01, 0.01$ and also $m_4/m_{\rm b} = 0.05, 0.1, 0.2$ and 0.3 for Model C1 which 
cover the range from planetary to stellar masses.
We again determine $i_{\rm max}$, the maximum inclinations of the outer object orbit relative to the
binary orbital plane using {\sc rebound} simulations. The simulations are run for $10^6 P_{\rm b}$.
These cases are then similar to the red points in the middle and bottom panel of Figure~\ref{fig:imax}
but with different outer object masses. 

Figure~\ref{fig:imaxmeb} plots results for the different cases. Note that the horizontal axis is 
$a_4$ normalized by $a_{\rm o}$,
the predicted outer semi-major axis of the unstable region that assumes $m_4=0$.
Since these models generally have $m_4>0$, the analytic curve in the
panels may not be accurate, especially at larger values of $m_4$.
The upper two panels show that the predictions of the analytic model 
work well for  $m_4 \le 0.01 m_{\rm b}$, which are typically in the planetary mass regime. 
For the case of $m_4 = 0.01 m_{\rm b}$, we see that inner and outer semi-major axes for the unstable region
shifts outward. This shift is larger at lower binary eccentricity.
The bottom panel Figure~\ref{fig:imaxmeb} shows that the instability is present even for stellar mass objects.
The outward shift of the unstable zone increases with increasing values of $m_4$.
The unstable zone of $m_4=0.3 m_{\rm b}$ lies  completely outside the predicted unstable region for $m_4=0$
plotted as the blue solid line,
The values of $i_{\rm max}$  generally decrease with increasing values of $m_4$ but remain
significant even at the highest value considered of $m_4 = 0.3 m_{\rm b}$,

We further explored the nature of the instability at higher planet masses by examining
the properties of the eccentricities and the inclinations relative to the invariable plane 
that is fixed in the inertial frame.
The inclination variations relative to the invariable plane then are the inclination
changes in the inertial frame.
The results are plotted in Figure~\ref{fig:HighMass} for Model C1 with $m_4=0 m_{\rm b}$ and $a_4=a_{\rm m}=14.4 a_{\rm b}$ in left panels 
and $m_4=0.3 m_{\rm b}$ and $a_4=1.6 a_{\rm o}=37.4 a_{\rm b}$ in the right panels.
Generally, the results are very different for these two different outer object masses.
The upper panels plot the inclinations relative to the invariable plane.
The lower panels plot the inclinations relative to the orbital plane of the binary and also plot the binary
eccentricity. The eccentricities of the inner planet and outer object that begin at zero remain small, 
less than 0.05 and 0.002, respectively, for both outer object masses.

For the case that $m_4=0$ the invariable plane nearly coincides with the
binary orbital plane because the binary contains nearly all of the system's angular momentum.
Consequently, the inclination of the binary orbit is nearly
zero relative to the invariable plane as shown in the upper left panel of Figure~\ref{fig:HighMass}.
In addition, the inclination evolution for the inner planet  are nearly the same in the 
upper and lower panels of Figure~\ref{fig:HighMass}.  The same also holds for the outer planet.
The binary eccentricity remains nearly constant.

The case that $m_4=0.3 m_{\rm b}$ is much different. The angular momentum of the outer
object dominates. Consequently, the inclination of the orbit of the 
outer object is smaller relative to the invariable plane than
the inclination of the binary orbit that varies considerably, as seen in the upper left panel of Figure~\ref{fig:HighMass}.
In addition, the inclination of the orbit of the inner planet relative to the invariable plane varies considerably.
However, in the lower panel of Figure~\ref{fig:HighMass} we see that the orbit of the inner planet remains nearly
polar relative to the binary at all times. The eccentricity of the binary undergoes significant variations. The 
inclination of the outer object relative to the binary is mainly due to inclination changes of the binary in the
inertial frame.

\section{Discussion}
\label{sec:disc}

The model presented describes an instability and resonance in the circumbinary case that in some ways is similar to what occurs   around a binary member in  Kozai-Lidov oscillations (KL)  in the case of small initial particle eccentricity \citep{Kozai1962,Lidov1962}.
In that case, there is a resonance in which the nodal precession frequency of the particle matches its apsidal precession frequency of the longitude of the periapsis. In the case analyzed here, there is a resonance for small initial inclination
in which the nodal precession frequency of the particle matches the apsidal precession frequency of the binary. In quadrupole order, there is an exponential growth
of the initially small eccentricity in the KL case \citep[e.g.,][]{Tremaine2014, Lubow2021} and exponential growth
of an initially small inclination in the case described here.

Both cases involve particle orbit evolution along a separatrix (path that separates librating from circulating orbits and contains a cusp near the origin) in a phase portrait:
$e \sin(\omega)$ versus $e \cos(\omega)$ in the KL case \citep[e.g.,][]{Lubow2021, Tremaine2023} and $i \sin(\phi)$ versus $i \cos(\phi)$ in the current case as seen in Figure~\ref{fig:phpl}. Since the separatrix is a closed
loop in both cases, they both involve secular oscillations. In both cases,
the tilt of the straight line on the right side of the cusp in the separatrix corresponds to the phase that is then nearly constant in time, giving rise to resonance.

In both cases, the resonance occurs over a range of parameters. 
In the KL case, there is a range of initial inclinations of the planet for which
eccentricity instability occurs.
 In the current case, there is a range of initial orbital radii of the outer planet for inclination instability that increases with binary eccentricity as seen in 
Equation (\ref{width}).
The range of inclinations for eccentricity growth in the KL case occurs through the range of possible apsidal phases 
$0 \leq \omega < 360^\circ$  for which $d \omega/dt = 0$. This occurs at nearly fixed inclination while particle eccentricity remains small but grows.
In the current case,
the range of radii for inclination growth occurs through the range of possible nodal phases $0 \leq \phi < 360^\circ$  for which $d \phi/dt = 0$. In
both cases, the stationary phase condition breaks down after a stage of exponential growth. In the KL case this breakdown  occurs once order unity eccentricity values are achieved, for inclinations well above the critical angle. In the case analyzed here,  this breakdown  occurs once order unity inclination values (in radians) are achieved, for $a_4$ intermediate between $a_{\rm i}$ and $a_{\rm o}$.
Following the breakdown, there in a change in sign of the growth rate. Consequently, oscillations occur in both cases. 

Apart from the similarity in growth in the two cases,
there is some similarity in the analytic form for the evolution of the nodal phase. In the KL case the effects of the
resonance cause a contribution to the evolution of the nodal phase of the form $\arctan{(k_1 \tanh{(k_2 t)})}$ 
for some constants in time $k_1$ and $k_2$ (see the last term on the RHS of equation (13) in \cite{Lubow2021}). In the current case, the nodal phase evolution has the same form if $\phi_0=0$ in Equation (\ref{phit}).

Resonance is possible in the case analyzed here because the polar inner planet causes the binary to undergo retrograde precession that can match the retrograde  nodal 
precession of the nearly coplanar outer planet. In the case of a nearly coplanar inner planet, the binary would undergo prograde precession and the inclination instability would not occur, unless the outer planet is on a retrograde orbit.

Large inclination oscillations have also been previously analyzed in the context of retrograde particle orbits external to the Sun-Jupiter system 
\citep{Zanardi2018}.  In this case, the binary consists of the Sun-Jupiter system and there is no inner planet
to cause apsidal precession of the binary. Instead the Sun-Jupiter system undergoes prograde precession due to GR effects.
A particle on an initially nearly coplanar retrograde orbit can undergo tilt oscillations due to the matching of the binary precession
frequency with the nodal precession rate of the particle \citep[e.g.,][]{Naoz2017,Lepp2022}.  

In the case that the outer object is of stellar mass,  the inclination evolution of the 
planet in the inertial frame ($i_3$ in upper right panel of Figure~\ref{fig:HighMass})  looks similar to what is expected to KL oscillations with high initial inclination \citep[e.g., upper panel in Figure~1 of][]{Lubow2021}. However, unlike KL oscillations, in this case $e_3$ undergoes small amplitude oscillations $0 \le e_3 < 0.05$ and these oscillations are of much shorter period than the inclination oscillations.

With $m_4=0.3 m_{\rm b}$, as seen in the lower right panel of 
Figure~\ref{fig:HighMass}, 
the inclination of the 
outer object relative to the binary orbit is maximum
when the binary eccentricity is minimum, as occurs 
in KL oscillations.
Unlike the KL case, the orbits of the binary and the outer object are initially coplanar
and the binary must have a nonzero eccentricity. The vertical component
of angular momentum of the binary (along the direction of the system angular momentum) is approximately constant in time, similar to the KL case.

\section{Summary}
\label{sec: sum}
We have have analyzed the orbital tilt stability of a system consisting
of an eccentric orbit binary star and two circumbinary objects. The inner circumbinary
object is a planet on a circular polar orbit about the binary, while the outer object is a planet or star that is initially on an orbit that is circular or eccentric and coplanar with respect to the binary. We find that due to the effects of the polar planet, a tilt instability occurs over a range
of orbital semi-major axis values of the outer object, even if it is of stellar mass.

 For the case that the outer object is a planet, we extended the recent results of  \cite{Childs2023} to analytically determine
the maximum inclination and initial time dependence of the inclination growth of the outer planet. %The binary orbit inclination remains nearly coplanar with the invariable plane. 
We found that the inclination growth occurs as an instability that
can be understood as the result of a resonance in which the nodal precession frequency of the outer planet matches the apsidal
precession frequency of the binary. The resonance condition is satisfied over a broad range of outer planet radii, provided that the binary
is sufficiently eccentric (see Equation (\ref{width})). Following a relatively short initial adjustment phase, the inclination
growth occurs exponentially in time at a rate that is comparable to the absolute value of apsidal precession rate of the binary
for moderate binary eccentricity. The growth rate increases with binary eccentricity (Figure~\ref{fig:grm}).
%The analytic predictions for the initial growth of inclination and the maximum inclination 
%agree well with the results of four-body simulations (Figure \ref{fig:imax}).
%, \ref{fig:REBOUNDinclinEvolcirc}, \ref{fig:eccpl}, and \ref{fig:octl}).
The inclination evolution can be understood in terms of trajectories in a phase portrait (see Figure \ref{fig:phpl}).
%The instability persists when the outer planet is on an eccentric orbit (see Figures \ref{fig:eccpl} and \ref{fig:octl}). 
We tested the model with {\sc rebound} simulations.
The values of the maximum inclination achieved during the oscillations agree well with our
analytic model, especially at high binary eccentricity, even if the outer planet is on an eccentric orbit  (see Figure~\ref{fig:imax}).
As the binary mass ratio becomes extreme, the unstable zone of the outer planet shifts inwards. For sufficiently extreme binary mass ratios, the model breaks down and the evolution is dominated by other effects, such as strong planet-planet interactions (Figures \ref{fig:q} and \ref{fig:q0p01i}).

If the outermost object is of stellar mass, the four bodies form a triple star system with
a planet in a polar orbit about the central binary. Due to the effects of the polar planet on the apsidal precession rate of the binary,
the binary orbit undergoes significant 
tilt oscillations for a range of orbital semi-major axis values of the outer star.
The planet remains in a polar orbit about the binary
as both it and the central binary undergo tilt oscillations (see right panels of Figure~\ref{fig:HighMass}).

%If the inner planet becomes too massive, the binary will undergo KL oscillations 
% and the process would operate differently. It would be useful to explore  such configurations. 

Nearly all the circumbinary planets that have been found were detected by the transit method using the
Kepler and TESS telescopes \citep[e.g.,][]{Welsh2018,Orosz2019,Kostov2020}. This method relies on repeated transiting events
of planets on orbits that are nearly coplanar with the binary orbital plane. The inclination instability would make the detection of planets in the instability zone very unlikely. 
Alternative methods of detection such as binary transit timing variataions 
and binary Doppler measurements can detect
noncoplanar circumbinary planets \citep{Zhang2019, Standing2023}.

It may be possible that initially coplanar gaseous circumbinary discs are subject to this instability,
provided they are sufficiently long-lived. %The instability could result
%in coherent tilt oscillations of the disc as can occur for KL unstable
%discs \citep{Martin2014a}. 
Part of the disc may lie within the unstable 
zone, while other parts that lie closer to the binary would be in the stable zone.
For a disc that behaves rigidly, the overall effect of an inner stable region may weaken and even suppress the instability. But if it does not behave rigidly, the unstable portions might undergo tilt oscillations resulting in disc warping.

Circumbinary debris discs that are initially coplanar with the binary orbital plane could be subject to this instability.
The solid bodies within the disc could 
undergo tilt oscillations whose properties are sensitive to their initial conditions. 
The instability could result in the 
vertical spreading of solid bodies that lie within the range of unstable semi-major axis values.
Objects that are close together could lie on different orbital planes 
and undergo strong collisions. The end result might be an apparent gap
in the disc.

%Due to the dissipation in the disc, it may evolve to a
%stationary inclination given by equation (8) of \cite{Childs2023}.

\section*{Acknowledgements}
AC acknowledge support from the NSF through grant NSF AST-2107738. RGM and SHL acknowledge support from NASA through grants 80NSSC19K0443 and 80NSSC21K0395. SHL thanks the Institute for Advanced Study for visitor support and thanks Scott Tremaine for a useful discussion.

%%%%%%%%%%%%%%%%%%%%%%%%%%%%%%%%%%%%%%%%%%%%%%%%%%
\section*{Data Availability}
 The $n$-body simulation results can be reproduced with the {\sc rebound} code (Astrophysics Source Code Library identifier {\tt ascl.net/1110.016}) and the {\sc reboundx} code (Astrophysics Source Code Library identifier {\tt ascl.net/2011.020}).  The data underlying this article will be shared on reasonable request to the corresponding author.



\bibliographystyle{mnras}
\bibliography{main} % if your bibtex file is called example.bib

%\begin{thebibliography}{99}
%\end{thebibliography}

%%%%%%%%%%%%%%%%%%%%%%%%%%%%%%%%%%%%%%%%%%%%%%%%%%

\appendix
\section{Transforming to the rotating frame}

The Hamiltonian in the nonrotating frame that is given by Equation (\ref{H}) can be 
 can be expressed in terms of  Delaunay canonical variables \citep[e.g.,][]{Tremaine2023} as
%\begin{equation}
%H_{\rm npc} = \alpha \left(\frac{\Lambda}{L} \right)^{3/2} \left[ (2+3e_{\rm b}^2) \left(1-3 \left(\frac{L_{\rm z}}{L} \right)^2  \right) - 15 e_{\rm b}^2 \cos{(2 \Omega_4)} \left(1-\left(\frac{L_{\rm z}}{L}\right)^2  \right)  \right] ,
 % \label{Hcan} 
%\end{equation}
\begin{equation}
H_{\rm npc} = \alpha \left(\frac{\Lambda}{L} \right)^{3} ( h_1 +h_2) ,
  \label{Hcan} 
\end{equation}
where $\alpha$ is defined in Equation (\ref{alpha}) and
\begin{eqnarray}
h_1 &=&  (2+3e_{\rm b}^2) \left(1-3 \left(\frac{L_{\rm z}}{L} \right)^2  \right), \label{h1} \\ 
h_2 &=&  - 15 e_{\rm b}^2 \cos{(2 \Omega_4)} \left(1-\left(\frac{L_{\rm z}}{L}\right)^2  \right), \label{h2} \\
\Lambda &=& a_{\rm b}^{3/2} \Omega_{\rm b} \sqrt{a_4},, \\
L &=&  a_{\rm b}^{3/2} \Omega_{\rm b} \sqrt{a_4 (1-e_4^2)}, \label{L} \\
L_z &=&  a_{\rm b}^{3/2} \Omega_{\rm b} \sqrt{a_4 (1-e_4^2)} \cos{i_4}.
\end{eqnarray}
The transformation to the rotating frame involves
changing the canonical coordinates in Equations (\ref{h1}) and (\ref{h2}) to the rotating frame (in this case just $\Omega_4$), while setting the canonical momenta to their values in the nonrotating frame.
Consequently, the longitude of the ascending node $\Omega_4$ in the nonrotating frame transforms to the longitude
of the ascending node in the rotating frame that we denote by $\phi_4$.
In addition,
the orbital inclination $i_4 = \arccos{(L_z/L })$, which is a function of the canonical momenta, does not change in transforming to the rotating frame, as expected.
The Hamltonian is also transformed  to
\begin{equation}
H = H_{\rm npc}- \Omega_{\rm f} L_z \label{Hr},
\end{equation} 
where $\Omega_{\rm f}$ is the rotation rate of the frame \citep{Tremaine2014, Tremaine2023}.
Expressing the Hamiltonian of Equation (\ref{Hr}) in terms of
 variables $\phi_4$ and $i_4$, we obtain Equation (\ref{Ht}) of the text.

We apply Hamilton's equations to Equation (\ref{Hr}) to obtain
\begin{equation}
\frac{d \Lambda}{dt} = -\frac{\partial H}{\partial \ell_4} =0,
\end{equation}
where  $\ell_4$ is the mean anomaly.
This equation implies that $a_4$ is constant in time, as is stated in Equation (\ref{dadt}) of the text.
We also have that
\begin{equation}
\frac{d L}{dt} = -\frac{\partial H}{\partial \omega_4} = 0,
\end{equation}
where $\omega_4$ is the argument of periapsis.
This equation and the constancy of $a_4$ imply that $e_4$ is constant in time, as is stated in Equation (\ref{dedt}) of the text.
Hamilton's equations also give that
\begin{equation}
\frac{d L_z}{dt} = -\frac{\partial H}{\partial \phi_4} 
\end{equation}
and
\begin{equation}
\frac{d \phi_4}{dt} = \frac{\partial H}{\partial L_z},
\end{equation}
which imply Equations (\ref{didt}) and (\ref{dphidt}) respectively  in the text.

% Don't change these lines
\bsp	% typesetting comment
\label{lastpage}
\end{document}